\documentclass[preprint,cleveref]{l4dc2027}

\title[High-Dimensional Implicit HJ Reachability]
{Implicit DeepReach Without Implicit Differentiation}
\usepackage{times}
\usepackage{enumitem}
\usepackage{cancel}
\usepackage{hyperref}
\usepackage{booktabs}
\usepackage{multirow}
\usepackage{multicol}
\usepackage{wrapfig}
\usepackage{algorithm2e}
\usepackage{makecell}
\usepackage{thmtools}

\usepackage[dvipsnames]{xcolor}
\author{%
 \Name{Xingjian Li} \Email{xingjian.li@austin.utexas.edu}\\
 \addr Oden Institute, The University of Texas at Austin
 \AND
 \Name{Kelvin Kan} \Email{kelvin.kan@math.ucla.edu}\\
 \addr Department of Mathematics, UCLA
 \AND
 \Name{Deepanshu Verma} \Email{dverma@clemson.edu}\\
 \addr School of Mathematical and Statistical Sciences, Clemson University
 \AND
 \Name{Krishna Kumar} \Email{krishnak@utexas.edu}\\
 \addr Oden Institute, The University of Texas at Austin
 \AND
 \Name{Hayden Schaeffer} \Email{hayden@math.ucla.edu}\\
 \addr Department of Mathematics, UCLA
 \AND
 \Name{Samy Wu Fung} \Email{swufung@mines.edu}\\
 \addr Department of Applied Mathematics and Statistics, Colorado School of Mines
}

\usepackage{xcolor}

\usepackage{amsmath,amsfonts,bm}

\def\1{\bm{1}}

\newcommand{\D}{\mathcal{D}}
\newcommand{\U}{\mathcal{U}}

\DeclareMathAlphabet{\mathsfit}{\encodingdefault}{\sfdefault}{m}{sl}
\SetMathAlphabet{\mathsfit}{bold}{\encodingdefault}{\sfdefault}{bx}{n}

\newcommand{\norm}[1]{\ensuremath{\left\|#1\right\|_2}}

\renewcommand{\eqref}[1]{\textup{(\ref{#1})}}

\begin{document}

\maketitle

\begin{abstract}
We consider the problem of approximating backward reachable and reach-avoid tubes for high-dimensional nonlinear systems whose Hamiltonians require numerical optimization over constrained controls and disturbances. 
Existing neural Hamilton–Jacobi (HJ) reachability methods alleviate the dimensionality bottleneck of grid-based solvers, but evaluating Hamiltonians without closed-form expressions introduces an additional computational challenge. 
We show that, under suitable assumptions, the gradients required for value-function training can be computed without differentiating through the inner optimization problem. Specifically, we establish an envelope-type gradient identity for the saddle point Hamiltonian, under which the optimal control and disturbance enter backpropagation as constants without gradient tracking. This enables end-to-end neural HJ reachability with numerically evaluated Hamiltonians while avoiding both solver unrolling and implicit differentiation. We demonstrate the accuracy and computational efficiency of the proposed approach on nonlinear reachability and reach-avoid problems of different scales.
\end{abstract}


\section{Introduction}

Hamilton--Jacobi (HJ) reachability provides a principled framework for reasoning about safety and robustness in dynamical systems. Given a target or unsafe set, reachability analysis characterizes the states from which an adversarial disturbance can force the system into the target despite the action of a controller \citep{mitchell2005,isaacs1965}. The resulting backward reachable tube (BRT) provides both a safety certificate and an associated feedback controller, making HJ reachability a powerful tool for safety-critical control.

A primary challenge in HJ reachability is the curse of dimensionality. Classical methods solve the associated Hamilton-Jacobi-Isaacs (HJI) equation or variational inequality on a state-time grid \citep{mitchell2005time,kundu2026policy}, these methods are accurate but restricted to relatively low-dimensional systems. 
Neural approaches such as DeepReach \citep{bansal2021deepreach} alleviate this bottleneck by representing the value function with a neural network and enforcing the HJI equation at sampled state-time points. Subsequent work has further improved the scalability and accuracy of this paradigm through additional supervision and guidance \citep{sharpless2025linear,teoh2025madr}.

In this work, we focus on a separate bottleneck that arises when the Hamiltonian of the system does not admit a closed form solution and is instead defined \emph{implicitly} by a saddle point problem. 
While optimal control and disturbance inputs can admit closed-form expressions in some standard reachability problems, more general nonlinear dynamics, coupled control–disturbance interactions, and complex admissible input sets generally require solving a numerical optimization problem to evaluate the Hamiltonian~\citep{betts2010practical,gelphman2025end,gelphman2026convergence}.
Recent approaches address such Hamiltonians through learned surrogates, data-driven approximations, or conservative bounds~\citep{chilakamarri2025blackbox,choi2025ddh,lin2026rover}. These methods make the Hamiltonian more tractable before or outside the value-function computation, but they usually require additional computation or data collection and can introduce approximation error.

An alternative is to retain the numerical Hamiltonian optimization directly within neural HJI training, as explored in related implicit deep learning approaches for optimal control \citep{gelphman2026convergence}. However, embedding the optimization into the training loop introduces an additional differentiation bottleneck. Each value-function update requires repeatedly solving the pointwise control-disturbance game, and conventional differentiation approaches propagate gradients through its solution. 
Unrolling the inner solver can incur substantial memory and computational overhead, while exact implicit differentiation typically requires solving an additional linear system for each Hamiltonian evaluation. In this work, we show that neither approach is necessary. By establishing an envelope-type gradient identity for the saddle point Hamiltonian, we obtain the gradients required for neural HJI training without differentiating through the inner optimization. 
In particular, our main contributions are:
\textbf{(i) End-to-end HJ reachability with implicit Hamiltonians.}
We develop a neural HJ reachability framework that directly incorporates numerically solved control-disturbance games into value-function training, enabling the treatment of nonlinear systems without tractable closed-form Hamiltonians.
\textbf{(ii) Theoretically justified gradients without optimizer sensitivities.}
Using a saddle point envelope argument, we establish exact gradient formulas under light assumptions, eliminating the need for solver unrolling or implicit differentiation through the inner optimization.
\textbf{(iii) Numerical validation.}
We demonstrate the accuracy and scalability of our approach on nonlinear reachability and reach-avoid problems in dimensions ranging from 2 to 50, showing close agreement with reference solutions and reliable closed-loop performance.

\section{Background}
\label{sec:background}

\subsection{Hamilton-Jacobi Reachability}

Consider the controlled dynamical system over a finite time horizon 
\begin{equation}
    \dot{x}(s)=f\bigl(x(s),u(s),d(s)\bigr),
    \quad s\in[t,T],
    \label{eq:dynamics}
\end{equation}
where $x(s)\in\mathbb{R}^n$ is the state, $u(s)\in\mathcal U$ is the control, and $d(s)\in\mathcal D$ is an adversarial disturbance with $t$ denoting an arbitrary initial time and $T$ is a fixed terminal time. 
Let the target set be represented by
\begin{equation}
    \mathcal L=\{x\in\mathbb{R}^n:\ell(x)\leq 0\},
\end{equation}
for a sufficiently regular target function $\ell$.
The target set can represent either a set of goal states or unsafe states, and $\ell(x)$ is typically chosen as a signed distance function.

The \textbf{Backward Reachable Tube (BRT)}~\citep{mitchell2005,bansal2021deepreach} of a given problem consists of states from which the disturbance can force the system into $\mathcal L$ at some time over $[t,T]$ despite the action of the controller. This is characterized via the value function 
\begin{equation}
    V(x,t) =\inf_{d(\cdot)}\sup_{u(\cdot)}\min_{s\in[t,T]}\ell\bigl(x(s)\bigr),
\label{eq:value}
\end{equation}
and the BRT is given by the zero sublevel set of the value function for any given $t$, that is,
\begin{equation}
    \mathcal V(t) = \{x\in\mathbb{R}^n:V(x,t)\leq0\}.
\label{eq:brt}
\end{equation}
Define the HJI Hamiltonian
\begin{equation}
    H(x,p) := \max_{u\in\mathcal U}\min_{d\in\mathcal D}p^\top f(x,u,d).
    \label{eq:hamiltonian}
\end{equation}
where $p$ is the adjoint variable to the dynamics. 
Under standard regularity assumptions and the Isaacs condition, the value function satisfies the HJI variational inequality (VI), which reads
\begin{equation}
    \min\left\{\partial_t V(x,t) + H\bigl(x,\nabla_xV(x,t)\bigr),\  \ell(x)-V(x,t) \right\} = 0,
    \label{eq:hji_vi}
\end{equation}
with terminal condition $V(x,T)=\ell(x)$ for all $x \in \mathbb{R}^n$. 
Crucially, the VI formulation accounts for the possibility of reaching the target set at any time within the horizon, rather than only at the terminal time, distinguishing it from the standard HJI PDE.

Similarly, we can define the \textbf{Backward Reach-Avoid Tube (BRAT)}~\citep{fisac2015reach}, which characterizes states from which the controller can reach the target set $\mathcal L$ despite adversarial disturbances, while avoiding an unsafe set $\mathcal G$, defined as
\begin{equation}
    \mathcal G=\{x\in\mathbb R^n:g(x)>0\}. 
\end{equation}
Here $g$ is also typically chosen as a signed distance function describing obstacles or unsafe states of the system, such as joint positions that violate a robot's safety requirements.
We note for goal-reaching problems, the roles of the control and disturbance in the Hamiltonian are reversed. Nonetheless, the BRAT can be similarly defined as the zero sublevel set of  the corresponding value function $V(x, t)$, which satisfies a slightly different variational inequality
\begin{equation}
\max\left\{ \min\left\{\partial_t V(x,t) + H(x,\nabla_x V(x,t)), \ \ell(x)-V(x,t)\right\}, \ g(x)-V(x,t) \right\}=0,
\label{eq:hji_vi_brat}
\end{equation}
with terminal condition $V(x,T)=\max\{\ell(x),g(x)\}$. Here the outer maximum enforces the state constraint by excluding trajectories that enter the unsafe set during the horizon. The BRAT $\mathcal  V(t)$, can be recovered similarly following~\eqref{eq:brt}.

\subsection{Neural HJ Reachability}

To alleviate the curse of dimensionality of grid-based HJ solvers, neural approaches such as DeepReach~\citep{bansal2021deepreach} approximate the value function using a neural network $V_\theta(x,t)$ parameterized by $\theta$. Following the physics-informed neural network (PINN) paradigm~\citep{raissi2019physics,wang2023expert}, the network is trained by minimizing the HJI variational inequality residual and enforcing the terminal condition at sampled state-time points. Specifically, the training objective takes the form
\begin{equation}
    {L}(\theta) = L_{\mathrm{VI}}(\theta) + \lambda L_{T}(\theta),
    \label{eq:neural_hj_loss}
\end{equation}
where
\begin{align}
    L_{\mathrm{VI}}(\theta) 
    & = \mathbb{E}_{(x,t)\sim\rho}\left[\left(\min\left\{\partial_t V_\theta(x,t) + H(x,\nabla_x V_\theta(x,t)), \ \ell(x)-V_\theta(x,t)\right\} \right)^2 \right],
    \label{eq:hji_residual_loss}
    \\
    L_{T}(\theta)
    & = \mathbb{E}_{x\sim\rho_T} \left[\left( V_\theta(x,T)-V(x,T) \right)^2\right].
    \label{eq:terminal_loss}
\end{align}
Here, $\rho$ and $\rho_T$ denote the sampling distributions over interior state-time points and terminal states, respectively, and $\lambda>0$ balances the two loss terms. 
The formulation above corresponds to the BRT problem; the BRAT loss follows analogously using the reach-avoid variational inequality.

A key requirement of these approaches is the repeated evaluation of the Hamiltonian $H(x,\nabla_x V_\theta)$ during training. 
Existing neural HJ methods typically consider settings in which the saddle point problem
admits a closed-form solution. However, for more complex dynamics, determining the optimal control and disturbance may require numerical optimization. Addressing the resulting differentiability and computational cost bottlenecks remains an open challenge for neural HJ methods.

\begin{wrapfigure}{t}{0.5\textwidth}
\vspace{-1em}
\begin{minipage}{\linewidth}
\setlength{\algomargin}{0pt}
\begin{algorithm2e}[H]
\DontPrintSemicolon
\SetAlgoLined
\KwIn{Dynamics $f$, target $\ell$, value network $V_\theta$, learning rate $\alpha$}
\KwOut{Trained value function $V_\theta$}

\While{not converged}{
    Sample $\{(x_i,t_i)\}_{i=1}^{N}\sim\rho$\;
    $p_i\gets \nabla_xV_\theta(x_i,t_i)$\;
    $(V_i, V_{t,i})\gets(V_\theta(x_i,t_i), \partial_tV_\theta(x_i,t_i))$\;

    \For{$i=1,\ldots,N$}{
        $(u_i^\star,d_i^\star)\gets
        \operatorname{SaddleSolve}
        (x_i,\textcolor{blue!70!black}{\operatorname{stopgrad}(p_i)})$\;
        $H_i\gets p_i^\top
        f(x_i,\textcolor{blue!70!black}{\operatorname{stopgrad}(u_i^\star,d_i^\star)})$\;
        $R_i\gets\min\{V_{t,i}+H_i,\ell(x_i)-V_i\}$\;
    }
    $L\gets\frac{1}{N}\sum_{i=1}^{N}R_i^2$\;
    $\theta\gets\theta-\alpha\nabla_\theta L$\;
}
\caption{Gradient-free neural HJ reachability  with implicit Hamiltonians.}
\label{alg:implicit_hj}
\end{algorithm2e}
\end{minipage}
\vspace{-2em}
\end{wrapfigure}

\section{Neural HJ Reachability with Implicit Hamiltonians}
\label{sec:method}

As discussed in Section~\ref{sec:background}, existing neural HJ reachability methods typically rely on Hamiltonians that admit closed-form evaluations. For general nonlinear dynamics with  constrained controls and disturbances, the Hamiltonian is defined implicitly through a saddle point optimization problem. Under a standard projected gradient ascent-descent (PGDA) formulation, the inner optimization can be recast as the fixed-point problem over $(u^\star, d^\star)$ with each iteration defined as
\begin{equation*}
\begin{aligned}
u^{k+1} &= \Pi_{\mathcal U}
\left(u^k + \eta_u \nabla_u
\left[p_\theta^\top f(x,u^k,d^k)\right]\right),\\
d^{k+1} &= \Pi_{\mathcal D}
\left(d^k - \eta_d \nabla_d
\left[p_\theta^\top f(x,u^k,d^k)\right]\right).
\end{aligned}
\end{equation*}
Here, $\Pi_{\mathcal U}$ and $\Pi_{\mathcal D}$ denote projections onto the admissible sets, and $\eta_u,\, \eta_d>0$ are the step sizes. 
Under suitable convex–concave assumptions PGDA can provably converge, more general saddle point problems may require alternative algorithms~\citep{korpelevich1976extragradient,letcher2019differentiable,mazumdar2025finding}.
Regardless of the solver, differentiating through the inner optimization introduces an additional computational bottleneck in neural HJI training.
Two conventional approaches exist: \textbf{(i) Automatic Differentiation (AD):} Unrolling the inner solver and backpropagating through its iterations, which can incur substantial computational and memory overhead especially as problem size and number of inner iterations increase. \textbf{(ii) Implicit Differentiation:} Using the implicit function theorem (IFT)~\citep{lorraine2020optimizing} it is possible to compute the sensitivities of the saddle point directly under
\begin{equation}
\begin{bmatrix}
\partial_\theta u^\star\\ \partial_\theta d^\star 
\end{bmatrix}
= -
\begin{bmatrix}
\nabla^2_{uu}(p_\theta^\top f) & \nabla^2_{ud}(p_\theta^\top f)\\ \nabla^2_{du}(p_\theta^\top f) & \nabla^2_{dd}(p_\theta^\top f)
\end{bmatrix}^{-1}
\begin{bmatrix}
\partial_\theta \nabla_u(p_\theta^\top f)\\ \partial_\theta \nabla_d(p_\theta^\top f)
\end{bmatrix}
\bigg|_{(u^\star,d^\star)}.
\label{eq:ift}
\end{equation}
Here we assume an interior saddle point for simplicity. While efficient, boundary projections can lead to singular or ill-conditioned Jacobians and subsequent numerical instabilities, especially in large-scale problems, as demonstrated in~\citep{li2026end}.
Both approaches introduce additional differentiation costs on top of the repeated saddle point solves required during training, rendering them impractical at scale. 

In this work, we show that differentiating through the inner optimization \emph{is unnecessary}. 
Leveraging a saddle point envelope type argument, we establish exact gradients for neural HJI training without differentiating through the inner solver, which has also been observed in the optimal transport setting~\cite{amos2023on, park2026implicit}. We state the formal theory in~\Cref{sec:theory} and present the resulting end-to-end training algorithm in~\Cref{sec:training_algorithm}.

\subsection{Theoretical Analysis}
\label{sec:theory}

We now establish an envelope type gradient identity that enables neural HJ reachability training without differentiating through the inner saddle point optimization.

\paragraph{Notation.} We write $z=(u,d)\in\U\times\D$, and use $z$ and $(u,d)$ interchangeably. Let $p_\theta:=\nabla_x V_\theta(x,t)$ and $\varphi(p,z):=p^\top f(x,z)$, so that $\displaystyle H(x,p)=\max_{u\in\U}\min_{d\in\D}\varphi(p,u,d)$ by~\eqref{eq:hamiltonian}.

\paragraph{Assumptions.} 
For an arbitrary parameter $\theta_0$, we assume: \textbf{(A1)} $\U$ and $\D$ are nonempty, compact and convex; \textbf{(A2)} $f(x,\cdot)$ is $C^1$; \textbf{(A3)} the mapping $\theta\mapsto p_\theta$ is differentiable at $\theta_0$; \textbf{(A4)} there exists $\delta_0=\delta(p_{\theta_0})>0$ such that, for every
$p \in \mathcal{B}_0 := \{ p \in \mathbb{R}^n: \norm{p-p_{\theta_0}}<\delta_0 \}$, the Isaacs condition
$\displaystyle\max_{u\in\mathcal{U}}\min_{d\in\mathcal{D}}\varphi(p,u,d)
  =\min_{d\in\mathcal{D}}\max_{u\in\mathcal{U}}\varphi(p,u,d)$
holds; \textbf{(A5)} the function $\varphi(p_{\theta_0},\cdot,\cdot)$ has a unique saddle point on $\mathcal{U}\times\mathcal{D}$, which we denote by $z^\star(p_{\theta_0})$.

\begin{restatable}{theorem}{mainthm}\label{thm:main}
    Under \textup{(A1)--(A5)}, the mapping $\theta\mapsto H(x,p_\theta)$ is differentiable at $\theta_0$, with
    \begin{align}
        \begin{split}
            \partial_\theta H(x,p_\theta)|_{\theta_0} &=f\bigl(x,z^\star(p_{\theta_0})\bigr)^\top\,(\partial_\theta p_\theta)|_{\theta_0} + \underbrace{\cancel{\bigl[\nabla_z\bigl(p_{\theta_0}^\top f(x,z)\bigr)\big|_{z=z^\star(p_{\theta_0})}\bigr]^\top \, \left(\partial_\theta z^\star(p_\theta)\right) |_{\theta_0}}}_{0 \ \text{contribution}}
            \\
            &=  f(x, z^\star(p_{\theta_0}))^\top \, (\partial_\theta p_\theta) |_{\theta_0}.
        \end{split}
    \end{align}
    In other words, the extra term 
    that would result from chain rule does not appear. Thus, the saddle point can be treated as a constant during backpropagation, with no gradients propagated through $z^\star(p_\theta)$ or the solver that computes it. 
\end{restatable}
We defer the proof of~\Cref{thm:main} to Appendix~\Cref{sec:proof}, numerical validation of~\Cref{thm:main} is also provided in Appendix~\Cref{sec:additional_results}.

\noindent\textbf{Remark.}
We highlight several important properties of \Cref{thm:main}.
\textbf{(i)}
Assumptions (A1)-(A5) are compatible with standard HJ reachability settings involving compact convex input constraints, sufficiently regular dynamics, and locally well-defined saddle points. In particular, the result does not require strong convexity-concavity or second-order nonsingularity conditions crucial for IFT.
\textbf{(ii)}
\Cref{thm:main} applies pointwise at any $\theta_0$ satisfying these assumptions and therefore extends throughout training wherever they continue to hold. We validate this numerically and measure the effect of inexact inner solves in Appendix~\Cref{sec:additional_results}, enabling our proposed~\Cref{alg:implicit_hj}.
\textbf{(iii)}
The result applies to arbitrary compact and convex control and disturbance sets, beyond simple box constraints. Moreover, the gradient identity is independent of the numerical saddle point solver.
\textbf{(iv)}
Finally, the result does not require differentiability of the optimal control and disturbance mappings with respect to the network parameters. Consequently, the gradient identity remains valid for saddle points in the interior or on the boundary of the admissible sets.

\subsection{Gradient-Free Training for Neural HJ Reachability with Implicit Hamiltonians}
\label{sec:training_algorithm}

Following the envelope identity established in~\Cref{sec:theory}, we develop an end-to-end neural HJ training framework that incorporates numerically evaluated Hamiltonians without differentiating through the inner saddle-point solver. We describe its main implementation components below.

\paragraph{Exact terminal condition enforcement through network parameterization.}
Following~\citep{singh2025exact}, we enforce the terminal condition $V_\theta(x,T)=\ell(x)$ by parameterizing the value function as
\begin{equation}
    V_\theta(x,t) = e^{-(T-t)}\ell(x) + \left(1-e^{-(T-t)}\right)N_\theta(x,t),
    \label{eq:hard_terminal}
\end{equation}
where $N_\theta$ denotes the neural network output. Unlike~\citep{singh2025exact}, we opt for an exponential decay to reduce the influence of the nonsmooth terminal function $\ell(x)$ away from $T$, which empirically we observe improves training. This parameterization satisfies the terminal condition exactly, eliminating the need for the terminal loss~\eqref{eq:terminal_loss}.

\paragraph{Saddle point solver.}
We primarily employ PGDA to solve the inner saddle point problem, using adaptive step sizes based on the Barzilai--Borwein (BB) method~\citep{barzilai1988two} to improve practical convergence. The step sizes are updated as
\begin{equation}
\begin{aligned}
\eta_u^{k+1} &= \operatorname{clip}\left( \frac{\|u^{k+1}-u^k\|_2^2} {(u^k-u^{k+1})^\top (\nabla_u \varphi^{k+1}-\nabla_u \varphi^k)}, \eta_{\min},\eta_{\max} \right),\\
\eta_d^{k+1} &= \operatorname{clip}\left( \frac{\|d^{k+1}-d^k\|_2^2} {(d^{k+1}-d^k)^\top
(\nabla_d \varphi^{k+1}-\nabla_d \varphi^k)}, \eta_{\min},\eta_{\max} \right).
\end{aligned}
\label{eq:bb_stepsize}
\end{equation}
Here, $\eta_{\min}$ and $\eta_{\max}$ are predefined bounds for numerical stability, and step sizes are updated independently for each sample. This provides a computationally inexpensive adaptive step-size selection, particularly useful when the saddle point objective is relatively flat. We evaluate its effectiveness against a fixed step size solver in~\Cref{sec:additional_results}. We note that while PGDA convergence depends on the structure of the underlying game, the proposed training framework is independent of the saddle point solver and can accommodate any alternative numerical methods.

Combining the exact terminal-condition parameterization with gradient-free Hamiltonian evaluation yields the end-to-end training procedure summarized in~\Cref{alg:implicit_hj}. After training, the BRT is recovered from the zero sublevel set of $V_\theta$, and the corresponding feedback policy can be obtained using the same saddle point solver. For conciseness, \Cref{alg:implicit_hj} is written in BRT formulation; the BRAT extension follows analogously.

\section{Experiments}
\label{sec:results}
We consider three numerical experiments, covering both BRT and BRAT problems, with state dimensions from $2$ to $50$. In all cases, the Hamiltonians are implicitly defined and require iterative solvers to compute the saddle points, and optimal saddle points occur both in the interior and on the boundaries of the admissible sets, providing sufficient numerical evidence supporting our approach. 
All experiments were conducted on a single NVIDIA RTX 5090 GPU. Code will be made available upon publication of the work. 
We refer to Appendix~\Cref{sec:example_details} and~\Cref{sec:training_details} for additional details.

\subsection{Energy Constrained Pursuit--Evasion Example}

We first consider a two-player zero-sum differential game with state $x=(r,q)\in\mathcal{X}=[0,2]\times[-0.5,1.5]$, where $r$ denotes the relative distance between the players and $q$ their relative energy resource, with $q>0$ indicating an advantage for the evader. The evader and pursuer select controls $u,d\in[0,2]$, respectively, under the dynamics
\begin{equation}
\begin{cases}
\dot r=(1+0.2q)\tanh(u)-1.2(1-0.1q)\tanh(d),\\
\dot q=-0.2u^2+0.2d^2.
\end{cases}
\end{equation}
The target set is defined by
$\mathcal{L}=\{(r,q):r\leq0.25\}\cup\{(r,q):q\leq0\}$,
with terminal function $\ell(r,q)=\min\{r-0.25,q\}$,  the target set is reached when either the relative distance falls below the given threshold or the evader exhausts its resource advantage.

Given the finite time horizon $T=1.5$, the BRT can be defined as in~\Cref{sec:background}, the Hamiltonian is separable in $u$ and $d$, allowing their optimal values to be computed independently from their respective stationarity conditions
\begin{equation}
\begin{cases}
p_r(1+0.2q)\operatorname{sech}^2(u)-0.4p_qu &= 0,\\
-1.2p_r(1-0.1q)\operatorname{sech}^2(d)+0.4p_qd &= 0,
\end{cases}
\end{equation}
subject to the input constraints, with $p = [p_r, p_q]^\top$ and $p=\nabla_x V(x,t)$. 
The Hamiltonian is provably convex-concave over the input domain, and the separable structure  enables comparison against a grid-based level-set solver while retaining nontrivial numerical optimization of the optimal inputs. 

\begin{figure}[t]
    \centering
    \includegraphics[width=0.95\linewidth]{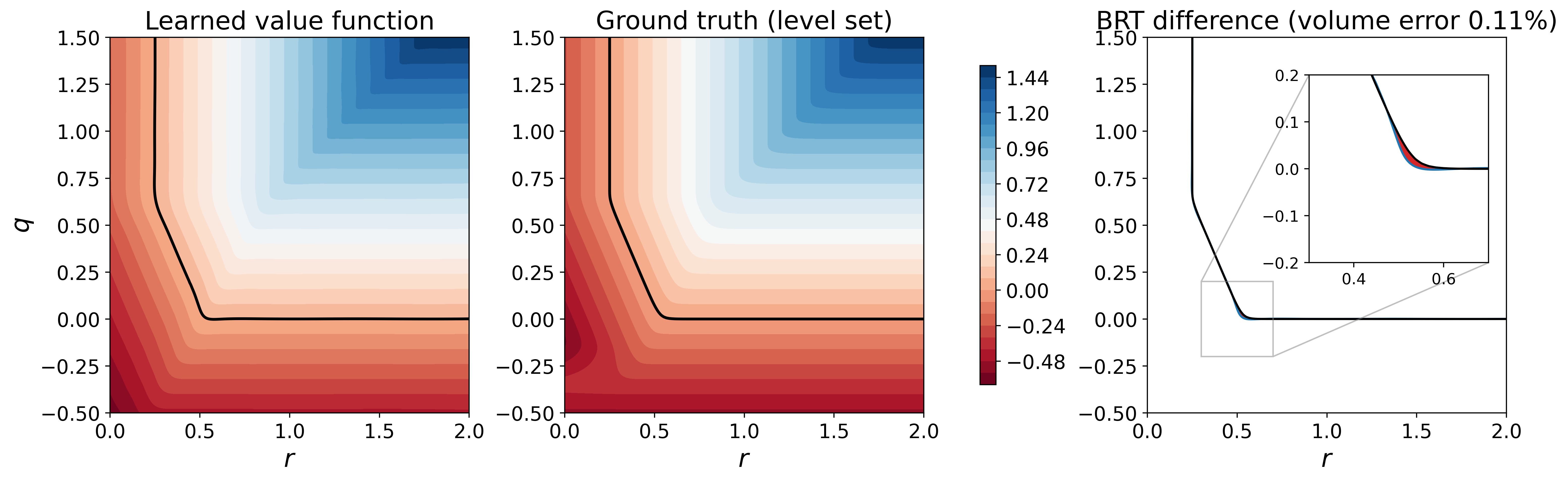}
    \caption{Learned and reference value functions for the energy-constrained pursuit-evasion problem. \textbf{Left:} Learned $V_\theta$; \textbf{Middle:} grid-based reference; \textbf{Right:} BRT volume differences.}
    \label{fig:energy-2d}
\end{figure}

We refer to Appendix~\Cref{sec:training_details} for details on model parameterization and training details. As shown in~\Cref{fig:energy-2d}, the learned value function closely matches the grid-based reference solution, achieving a training loss on the order of $10^{-6}$ and a BRT volume error of only $0.11\%$.

\subsection{Extending to High Dimensions}

We extend the 2D energy-constrained pursuit-evasion example to $N$ independent evader-pursuer pairs, with joint state $x=(r_1,q_1,\ldots,r_N,q_N)\in\mathbb{R}^{2N}$ and dynamics
\begin{equation}
\begin{cases}
\dot r_i=(1+k_i^1q_i)\tanh(u_i) -1.2(1-k_i^2q_i)\tanh(d_i),\\
\dot q_i=-\alpha_i u_i^2+\beta_i d_i^2,
\end{cases}
\quad i=1,\ldots,N.
\end{equation}
The target set is defined by $\mathcal{L}=\{x:\ell(x)\leq 0\}$, where $\ell(x)=\min_{i=1,\ldots,N} \{r_i-r_i^{\mathrm{reach}},q_i-q_i^{\mathrm{reach}}\}$ so that the target is reached whenever any pair violates its separation or energy threshold..
We refer to Appendix~\Cref{sec:example_details} for the complete problem formulation.

\begin{figure}[t]
    \centering
    \includegraphics[width=0.9\linewidth]{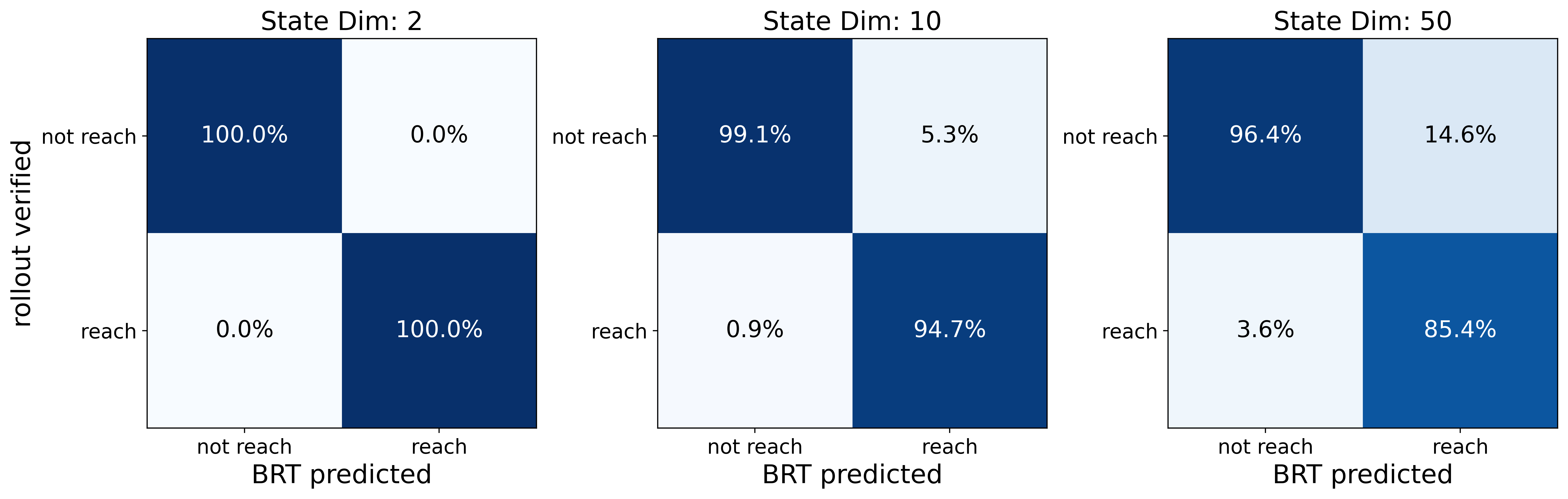}
    \caption{Closed-loop consistency of learned BRT predictions across state dimensions, quantifying mismatches between BRT predictions and closed-loop rollouts. Confusion matrices are column-normalized by predicted class.}
    \label{fig:confusion-matrix}
\end{figure}

We consider 10D and 50D extensions to evaluate scalability of our proposed method.
As shown in \Cref{fig:scalability-results}, training converges in both cases, with increasing residual losses at higher dimensions given the added complexity.
To assess the learned BRTs without grid-based reference solutions,  we perform a closed-loop consistency test on $2048$ randomly sampled initial states for each problem, comparing BRT
predictions with rollouts under the learned saddle-point policy, similar to that of~\citep{bansal2021deepreach}.
The column-normalized confusion matrices in~\Cref{fig:confusion-matrix} show that agreement remains high, particularly for states predicted to be unreachable ($99.1\%$
in 10D and $96.4\%$ in 50D). Agreement for predicted-reachable states decreases to $94.7\%$ and $85.4\%$, respectively, indicating greater classification disagreement as dimensionality increases.

\subsection{Two-Vehicle Collision Avoidance in 6D}
We consider a two-vehicle collision avoidance problem based on~\citet{bansal2021deepreach}, with joint state $x=(x_1,\ldots,x_6)\in\mathbb{R}^4\times\mathbb{S}^1 \times\mathbb{S}^1$. Unlike the standard constant speed setting, we introduce steering-dependent speed for the evader, resulting in the modified dynamics
\begin{equation*}
\left\{
\begin{aligned}
\dot{x}_1 &= 0.75\cos x_3, &
\dot{x}_4 &= 0.75e^{-0.4(u/3)^2}\cos x_6,\\
\dot{x}_2 &= 0.75\sin x_3, &
\dot{x}_5 &= 0.75e^{-0.4(u/3)^2}\sin x_6,\\
\dot{x}_3 &= d, &
\dot{x}_6 &= u,
\end{aligned}
\right.
\label{eq:vehicle_dynamics}
\end{equation*}
where $u,d\in[-3,3]$ denote the evader's and pursuer's turn-rate controls, respectively. This modification introduces nonlinear control dependence into the Hamiltonian, requiring numerical optimization of the evader's optimal control. We follow the capture target formulation of
\citet{bansal2021deepreach}, with the complete problem definition given in Appendix~\Cref{sec:example_details}.

\begin{figure}[t]
    \centering
    \includegraphics[width=0.95\linewidth]{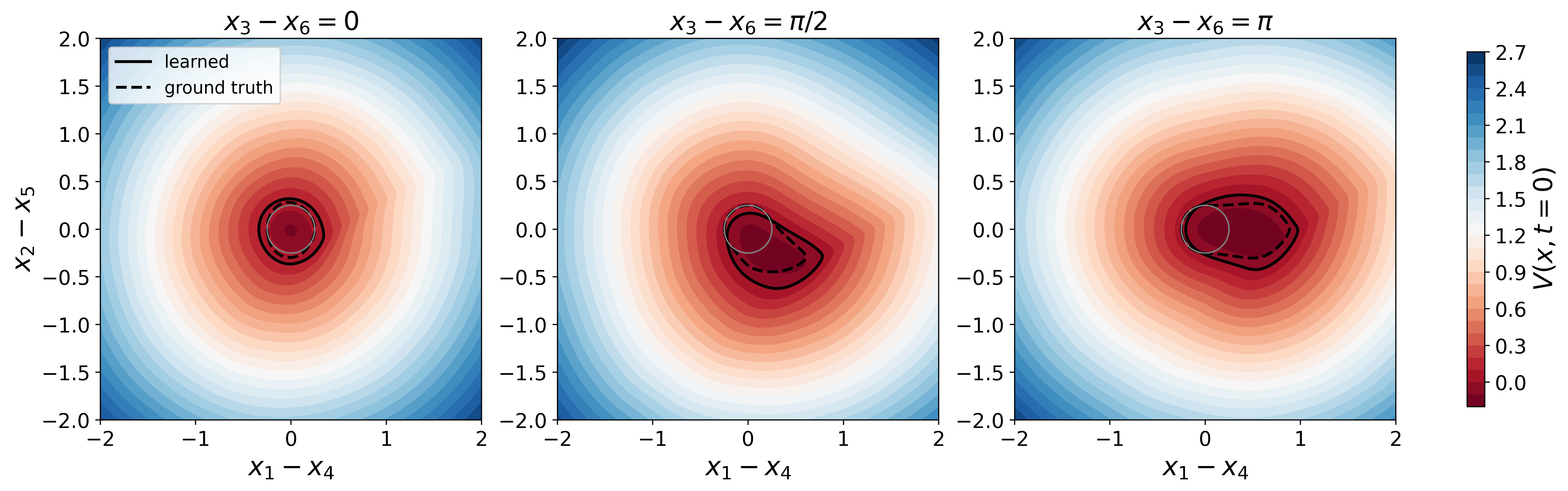}
    \caption{Learned and reference BRTs across relative-heading slices, showing close agreement with a grid-based reference solution.}
    \label{fig:two-vehicle-brt}
\end{figure}

Training yielded a MSE loss of approximately $2\times 10^{-3}$. 
We note although the model is trained directly in the full 6D state space, a reference solution can be computed in 3D using relative coordinates, exploiting the translation and rotation symmetries of the dynamics. 
As shown in \Cref{fig:two-vehicle-brt}, the learned BRT closely matches a grid-based reference solution across multiple two-dimensional slices, with slightly conservative predictions. 
Importantly, we obtain the desired results without enforcing said symmetries during neural HJ training.

We further evaluate the learned value function through closed-loop trajectory rollouts in
\Cref{fig:two-vehicle-traj} in the Appendix, using the same numerical Hamiltonian optimization as during training. 
The trajectories demonstrate both successful evasion and capture, with the inner solver achieving residual errors below $10^{-4}$ across the tested scenarios. The steering-dependent speed also changes the reachable set compared with the constant speed setting: by reducing its speed during rapid turning, the evader can escape from certain configurations that would otherwise lead to capture. The trajectories also illustrate how initially reachable states may leave the BRT over time.

\subsection{Quadrotor Reach-Avoid Control under Adversarial Disturbances}

We consider a 6D planar quadrotor reach-avoid problem under adversarial wind disturbances. The quadrotor is controlled through two bounded rotor thrusts and must reach a prescribed target region while avoiding ground collision and excessive pitch angles. The wind disturbance enters through nonlinear velocity-dependent drag, requiring numerical optimization to determine the worst case disturbance. We refer to Appendix~\Cref{sec:example_details} for the complete dynamics and problem formulation.

\begin{figure}
    \centering
    \includegraphics[width=1.0\linewidth]{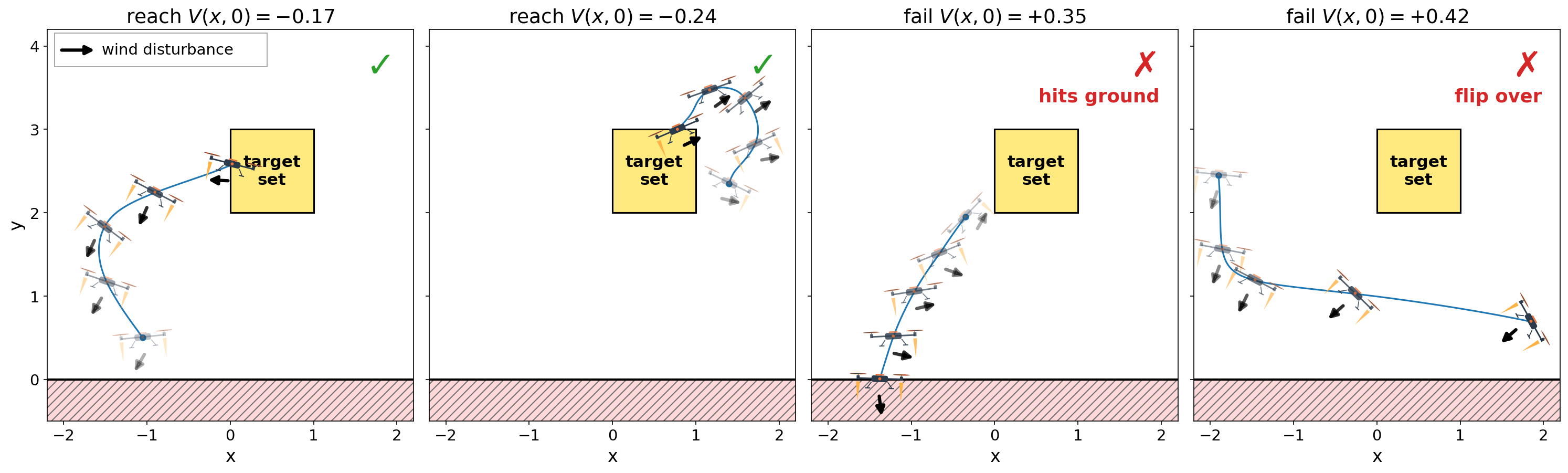}
    \caption{Rollouts for quadrotor control under disturbances using the learned $V_\theta$. Successful and failed trajectories are shown, with outcomes consistent with the BRAT predictions.}
    \label{fig:brat_rollouts}
\end{figure}

To address training difficulties arising from the nested BRAT variational inequality, we adopt a curriculum learning strategy common in PINNs literature~\citep{bansal2021deepreach,krishnapriyan2021characterizing,wang2022respecting}, where we gradually extend the sampled time horizon from $T$ to $0$ during training.
We evaluate the learned value function through a closed-loop consistency test on $2048$ randomly sampled initial states at $t=0$. As shown in~\Cref{tab:brat_consistency}, the rollout outcomes agree with the predicted reach-avoid classifications in over $90\%$ of cases for both predicted classes, with most discrepancies occurring near the zero level set. We further visualize representative trajectories in~\Cref{fig:brat_rollouts}, where feedback controls and worst-case disturbances derived from the learned value function produce outcomes consistent with the predicted BRAT.

\begin{table}[t]
\small
\centering
\caption{Closed-loop consistency of BRAT predictions using learned value function. ``Boundary'' denotes $|V_\theta(x,0)|<0.05$. Both crashed and timeout cases have $V_\theta(x,0) > 0$. }
\label{tab:brat_consistency}
\begin{tabular}{llrrrr}
\toprule
 & & Reached & Crashed & Timeout & Total \\
\midrule
\multirow{3}{*}{$V_\theta(x, 0) \le 0$}
 & \textbf{All} & \textbf{837 (90.8\%)} & \textbf{71 (7.7\%)} & \textbf{14 (1.5\%)} & \textbf{922} \\
 & Boundary      & 46           & 44           & 7            & 97  \\
 & Confident  & 791          & 27           & 7            & 825  \\
\midrule
\multirow{3}{*}{$V_\theta(x, 0) > 0$}
 & \textbf{All} & \textbf{62 (5.5\%)} & \textbf{911 (80.9\%)} & \textbf{153 (13.6\%)} & \textbf{1126} \\
 & Boundary      & 24           & 53  &  4      & 81  \\
 & Confident & 38           & 858  & 149     & 1045  \\
\bottomrule
\end{tabular}
\end{table}

\section{Conclusion}
We presented an end-to-end neural HJ reachability framework for nonlinear systems with implicitly defined Hamiltonians. 
Using a perturbation argument based on saddle-point inequalities, we prove that gradients can be computed without differentiating through the inner optimization, avoiding the overhead and numerical instability under solver unrolling or implicit differentiation. 
Experiments on both reachability and reach-avoid problems demonstrate the effectiveness and scalability of our approach, extending neural HJ methods to a much broader and more practical class of problems.

\clearpage

\bibliography{l4dc/references}

\begin{thebibliography}{45}
\providecommand{\natexlab}[1]{#1}
\providecommand{\url}[1]{\texttt{#1}}
\expandafter\ifx\csname urlstyle\endcsname\relax
  \providecommand{\doi}[1]{doi: #1}\else
  \providecommand{\doi}{doi: \begingroup \urlstyle{rm}\Url}\fi

\bibitem[Amos(2023)]{amos2023on}
Brandon Amos.
\newblock On amortizing convex conjugates for optimal transport.
\newblock In \emph{The Eleventh International Conference on Learning Representations}, 2023.
\newblock URL \url{https://openreview.net/forum?id=TQ5WUwS_4ai}.

\bibitem[Bansal and Tomlin(2021)]{bansal2021deepreach}
Somil Bansal and Claire~J. Tomlin.
\newblock {DeepReach}: A deep learning approach to high-dimensional reachability.
\newblock In \emph{Proceedings of the IEEE International Conference on Robotics and Automation (ICRA)}, pages 1817--1824. IEEE, 2021.
\newblock \doi{10.1109/ICRA48506.2021.9561949}.

\bibitem[Bartle and Sherbert(2000)]{bartle2000introduction}
Robert~G Bartle and Donald~R Sherbert.
\newblock \emph{Introduction to real analysis}, volume~2.
\newblock Wiley New York, 2000.

\bibitem[Barzilai and Borwein(1988)]{barzilai1988two}
Jonathan Barzilai and Jonathan~M Borwein.
\newblock Two-point step size gradient methods.
\newblock \emph{IMA journal of numerical analysis}, 8\penalty0 (1):\penalty0 141--148, 1988.

\bibitem[Betts(2010)]{betts2010practical}
John~T. Betts.
\newblock \emph{Practical Methods for Optimal Control and Estimation Using Nonlinear Programming}.
\newblock SIAM, 2010.

\bibitem[Chilakamarri et~al.(2025)Chilakamarri, Feng, and Bansal]{chilakamarri2025blackbox}
Vamsi~Krishna Chilakamarri, Zeyuan Feng, and Somil Bansal.
\newblock Reachability analysis for black-box dynamical systems.
\newblock In \emph{2025 IEEE International Conference on Robotics and Automation (ICRA)}, pages 3552--3558. IEEE, 2025.
\newblock \doi{10.1109/ICRA55743.2025.11127528}.

\bibitem[Choi et~al.(2025)Choi, Strong, Sreenath, Cho, and Tomlin]{choi2025ddh}
Jason~J. Choi, Christopher~A. Strong, Koushil Sreenath, Namhoon Cho, and Claire~J. Tomlin.
\newblock Data-driven hamiltonian for direct construction of safe set from trajectory data.
\newblock In \emph{2025 IEEE 64th Conference on Decision and Control (CDC)}, pages 2951--2958. IEEE, 2025.
\newblock \doi{10.1109/CDC57313.2025.11312763}.

\bibitem[Feng et~al.(2025)Feng, Qiu, and Bansal]{FengZ-RSS-25}
Zeyuan Feng, Le~Qiu, and Somil Bansal.
\newblock {Bridging Model Predictive Control and Deep Learning for Scalable Reachability Analysis}.
\newblock In \emph{Proceedings of Robotics: Science and Systems}, LosAngeles, CA, USA, June 2025.
\newblock \doi{10.15607/RSS.2025.XXI.114}.

\bibitem[Fisac et~al.(2015)Fisac, Chen, Tomlin, and Sastry]{fisac2015reach}
Jaime~F Fisac, Mo~Chen, Claire~J Tomlin, and S~Shankar Sastry.
\newblock Reach-avoid problems with time-varying dynamics, targets and constraints.
\newblock In \emph{Proceedings of the 18th international conference on hybrid systems: computation and control}, pages 11--20, 2015.

\bibitem[Fung et~al.(2022)Fung, Heaton, Li, McKenzie, Osher, and Yin]{fung2022jfb}
Samy~Wu Fung, Howard Heaton, Qiuwei Li, Daniel McKenzie, Stanley Osher, and Wotao Yin.
\newblock {JFB}: Jacobian-free backpropagation for implicit networks.
\newblock In \emph{Proceedings of the AAAI Conference on Artificial Intelligence}, volume~36, pages 6648--6656, 2022.

\bibitem[Gelphman et~al.(2026{\natexlab{a}})Gelphman, Verma, Yang, Osher, and Fung]{gelphman2025end}
Eric Gelphman, Deepanshu Verma, Nicole~Tianjiao Yang, Stanley Osher, and Samy~Wu Fung.
\newblock End-to-end training of high-dimensional optimal control with implicit hamiltonians via jacobian-free backpropagation.
\newblock In \emph{2026 American Control Conference (ACC)}. IEEE, 2026{\natexlab{a}}.

\bibitem[Gelphman et~al.(2026{\natexlab{b}})Gelphman, Verma, Yang, Osher, and Fung]{gelphman2026convergence}
Eric Gelphman, Deepanshu Verma, Nicole~Tianjiao Yang, Stanley Osher, and Samy~Wu Fung.
\newblock On the convergence of jacobian-free backpropagation for optimal control problems with implicit hamiltonians.
\newblock \emph{arXiv preprint arXiv:2602.00921}, 2026{\natexlab{b}}.

\bibitem[Hattenberger et~al.(2023)Hattenberger, Bronz, and Condomines]{hattenberger2023evaluation}
Gautier Hattenberger, Murat Bronz, and Jean-Philippe Condomines.
\newblock Evaluation of drag coefficient for a quadrotor model.
\newblock \emph{International Journal of Micro Air Vehicles}, 15:\penalty0 17568293221148378, 2023.

\bibitem[Heaton and Fung(2023)]{heaton2023explainable}
Howard Heaton and Samy~Wu Fung.
\newblock Explainable ai via learning to optimize.
\newblock \emph{Scientific Reports}, 13\penalty0 (1):\penalty0 10103, 2023.

\bibitem[Heaton et~al.(2021)Heaton, Wu~Fung, Gibali, and Yin]{heaton2021feasibility}
Howard Heaton, Samy Wu~Fung, Aviv Gibali, and Wotao Yin.
\newblock Feasibility-based fixed point networks.
\newblock \emph{Fixed Point Theory and Algorithms for Sciences and Engineering}, 2021\penalty0 (1):\penalty0 21, 2021.

\bibitem[Hofgard(2024)]{hofgard2024convergence}
William Hofgard.
\newblock Convergence guarantees for neural network-based hamilton{\textendash}jacobi reachability.
\newblock In \emph{NeurIPS 2024 Workshop on Data-driven and Differentiable Simulations, Surrogates, and Solvers}, 2024.
\newblock URL \url{https://openreview.net/forum?id=SH1isYejz9}.

\bibitem[Isaacs(1965)]{isaacs1965}
Rufus Isaacs.
\newblock \emph{Differential Games: A Mathematical Theory with Applications to Warfare and Pursuit, Control and Optimization}.
\newblock John Wiley \& Sons, New York, 1965.

\bibitem[Kingma and Ba(2014)]{kingma2014adam}
Diederik~P Kingma and Jimmy Ba.
\newblock Adam: A method for stochastic optimization.
\newblock \emph{arXiv preprint arXiv:1412.6980}, 2014.

\bibitem[Korpelevich(1976)]{korpelevich1976extragradient}
Galina~M Korpelevich.
\newblock The extragradient method for finding saddle points and other problems.
\newblock \emph{Matecon}, 12:\penalty0 747--756, 1976.

\bibitem[Krishnapriyan et~al.(2021)Krishnapriyan, Gholami, Zhe, Kirby, and Mahoney]{krishnapriyan2021characterizing}
Aditi Krishnapriyan, Amir Gholami, Shandian Zhe, Robert Kirby, and Michael Mahoney.
\newblock Characterizing possible failure modes in physics-informed neural networks.
\newblock \emph{Advances in neural information processing systems}, 34:\penalty0 26548--26560, 2021.

\bibitem[Kundu and Kunisch(2026)]{kundu2026policy}
Sudeep Kundu and Karl Kunisch.
\newblock Policy iteration for hamilton-jacobi-isaacs equations with control constraints and comparison with hamilton-jacobi-bellman equations.
\newblock \emph{arXiv preprint arXiv:2609.29368}, 2026.

\bibitem[Kwon and Eun(2024)]{kwon2024quadrotor}
Minhyeok Kwon and Yongsoon Eun.
\newblock Quadrotor dynamics in a wind field: Equilibria analysis and energy dissipation.
\newblock \emph{International Journal of Control, Automation and Systems}, 22\penalty0 (11):\penalty0 3275--3284, 2024.

\bibitem[Letcher et~al.(2019)Letcher, Balduzzi, Racaniere, Martens, Foerster, Tuyls, and Graepel]{letcher2019differentiable}
Alistair Letcher, David Balduzzi, S{\'e}bastien Racaniere, James Martens, Jakob Foerster, Karl Tuyls, and Thore Graepel.
\newblock Differentiable game mechanics.
\newblock \emph{Journal of Machine Learning Research}, 20\penalty0 (84):\penalty0 1--40, 2019.

\bibitem[Li et~al.(2026{\natexlab{a}})Li, Kan, Verma, Kumar, Osher, and Drgo{\v{n}}a]{li2026zero}
Xingjian Li, Kelvin Kan, Deepanshu Verma, Krishna Kumar, Stanley Osher, and J{\'a}n Drgo{\v{n}}a.
\newblock Zero-shot transferable solution method for parametric optimal control problems.
\newblock In \emph{2026 American Control Conference (ACC)}, pages 2003--2010. IEEE, 2026{\natexlab{a}}.

\bibitem[Li et~al.(2026{\natexlab{b}})Li, Kan, Verma, Kumar, Osher, and Fung]{li2026end}
Xingjian Li, Kelvin Kan, Deepanshu Verma, Krishna Kumar, Stanley Osher, and Samy~Wu Fung.
\newblock End-to-end learning of safe optimal feedback control in high dimensions with control barrier function layers.
\newblock \emph{arXiv preprint arXiv:2607.20674}, 2026{\natexlab{b}}.

\bibitem[Lin and Bansal(2023)]{lin2023generating}
Albert Lin and Somil Bansal.
\newblock Generating formal safety assurances for high-dimensional reachability.
\newblock In \emph{2023 IEEE International Conference on Robotics and Automation (ICRA)}, pages 10525--10531. IEEE, 2023.

\bibitem[Lin et~al.(2026)Lin, Pinto, and Bansal]{lin2026rover}
Albert Lin, Alessandro Pinto, and Somil Bansal.
\newblock Robust verification of controllers under state uncertainty via hamilton--jacobi reachability analysis.
\newblock In \emph{Proceedings of the 8th Annual Learning for Dynamics and Control Conference}, volume 331 of \emph{Proceedings of Machine Learning Research}, pages 1432--1446. PMLR, 2026.

\bibitem[Lorraine et~al.(2020)Lorraine, Vicol, and Duvenaud]{lorraine2020optimizing}
Jonathan Lorraine, Paul Vicol, and David Duvenaud.
\newblock Optimizing millions of hyperparameters by implicit differentiation.
\newblock In \emph{International conference on artificial intelligence and statistics}, pages 1540--1552. PMLR, 2020.

\bibitem[Mahony et~al.(2012)Mahony, Kumar, and Corke]{mahony2012multirotor}
Robert Mahony, Vijay Kumar, and Peter Corke.
\newblock Multirotor aerial vehicles: Modeling, estimation, and control of quadrotor.
\newblock \emph{IEEE robotics \& automation magazine}, 19\penalty0 (3):\penalty0 20--32, 2012.

\bibitem[Mazumdar et~al.(2025)Mazumdar, Sastry, and Jordan]{mazumdar2025finding}
Eric Mazumdar, S~Shankar Sastry, and Michael~I Jordan.
\newblock On finding local nash equilibria (and only local nash equilibria) in zero-sum games.
\newblock \emph{ACM/IMS Journal of Data Science}, 2\penalty0 (2):\penalty0 1--26, 2025.

\bibitem[McKenzie et~al.(2024{\natexlab{a}})McKenzie, Heaton, and Fung]{mckenzie2024differentiating}
Daniel McKenzie, Howard Heaton, and Samy~Wu Fung.
\newblock Differentiating through integer linear programs with quadratic regularization and davis-yin splitting.
\newblock \emph{Transactions on Machine Learning Research}, 2024{\natexlab{a}}.
\newblock ISSN 2835-8856.
\newblock URL \url{https://openreview.net/forum?id=H8IaxrANWl}.

\bibitem[McKenzie et~al.(2024{\natexlab{b}})McKenzie, Heaton, Li, Wu~Fung, Osher, and Yin]{mckenzie2024three}
Daniel McKenzie, Howard Heaton, Qiuwei Li, Samy Wu~Fung, Stanley Osher, and Wotao Yin.
\newblock Three-operator splitting for learning to predict equilibria in convex games.
\newblock \emph{SIAM Journal on Mathematics of Data Science}, 6\penalty0 (3):\penalty0 627--648, 2024{\natexlab{b}}.
\newblock \doi{10.1137/22M1544531}.

\bibitem[Mitchell et~al.(2005{\natexlab{a}})Mitchell, Bayen, and Tomlin]{mitchell2005}
Ian~M. Mitchell, Alexandre~M. Bayen, and Claire~J. Tomlin.
\newblock A time-dependent hamilton--jacobi formulation of reachable sets for continuous dynamic games.
\newblock \emph{IEEE Transactions on Automatic Control}, 50\penalty0 (7):\penalty0 947--957, 2005{\natexlab{a}}.
\newblock \doi{10.1109/TAC.2005.851439}.

\bibitem[Mitchell et~al.(2005{\natexlab{b}})Mitchell, Bayen, and Tomlin]{mitchell2005time}
Ian~M Mitchell, Alexandre~M Bayen, and Claire~J Tomlin.
\newblock A time-dependent hamilton-jacobi formulation of reachable sets for continuous dynamic games.
\newblock \emph{IEEE Transactions on automatic control}, 50\penalty0 (7):\penalty0 947--957, 2005{\natexlab{b}}.

\bibitem[Park et~al.(2026)Park, Gelphman, Osher, and Fung]{park2026implicit}
Yesom Park, Eric Gelphman, Stanley Osher, and Samy~Wu Fung.
\newblock Implicit neural optimal transport via fixed-point optimization.
\newblock \emph{arXiv preprint arXiv:2605.10792}, 2026.

\bibitem[Paszke et~al.(2019)Paszke, Gross, Massa, Lerer, Bradbury, Chanan, Killeen, Lin, Gimelshein, Antiga, et~al.]{paszke2019pytorch}
Adam Paszke, Sam Gross, Francisco Massa, Adam Lerer, James Bradbury, Gregory Chanan, Trevor Killeen, Zeming Lin, Natalia Gimelshein, Luca Antiga, et~al.
\newblock Pytorch: An imperative style, high-performance deep learning library.
\newblock \emph{Advances in neural information processing systems}, 32, 2019.

\bibitem[Raissi et~al.(2019)Raissi, Perdikaris, and Karniadakis]{raissi2019physics}
Maziar Raissi, Paris Perdikaris, and George~E Karniadakis.
\newblock Physics-informed neural networks: A deep learning framework for solving forward and inverse problems involving nonlinear partial differential equations.
\newblock \emph{Journal of Computational physics}, 378:\penalty0 686--707, 2019.

\bibitem[Schaeffer(2026)]{schaeffer2026neural}
Hayden Schaeffer.
\newblock Neural network approximation and structural complexity of semilinear pdes with lower bounds.
\newblock \emph{arXiv preprint arXiv:2610.40384}, 2026.

\bibitem[SeWell et~al.(2026)SeWell, Li, Tretiakov, Kumar, and Fridovich-Keil]{sewell2026neural}
David SeWell, Xingjian Li, Stepan Tretiakov, Krishna Kumar, and David Fridovich-Keil.
\newblock Neural operators for multi-task control and adaptation.
\newblock \emph{Transactions on Machine Learning Research}, 2026.
\newblock ISSN 2835-8856.
\newblock URL \url{https://openreview.net/forum?id=jciOb0z5Wm}.

\bibitem[Sharpless et~al.(2025)Sharpless, Feng, Bansal, and Herbert]{sharpless2025linear}
William Sharpless, Zeyuan Feng, Somil Bansal, and Sylvia Herbert.
\newblock Linear supervision for nonlinear, high-dimensional neural control and differential games.
\newblock In \emph{Proceedings of the 7th Annual Learning for Dynamics and Control Conference}, volume 283 of \emph{Proceedings of Machine Learning Research}, pages 365--377. PMLR, 2025.

\bibitem[Singh et~al.(2025)Singh, Feng, and Bansal]{singh2025exact}
Aditya Singh, Zeyuan Feng, and Somil Bansal.
\newblock Exact imposition of safety boundary conditions in neural reachable tubes.
\newblock In \emph{2025 IEEE International Conference on Robotics and Automation (ICRA)}, pages 5489--5495. IEEE, 2025.

\bibitem[Teoh et~al.(2025)Teoh, Tonkens, Sharpless, Yang, Feng, Bansal, and Herbert]{teoh2025madr}
Ryan Teoh, Sander Tonkens, William Sharpless, Aijia Yang, Zeyuan Feng, Somil Bansal, and Sylvia Herbert.
\newblock {MADR}: {MPC}-guided adversarial {DeepReach}.
\newblock \emph{arXiv preprint arXiv:2510.18845}, 2025.

\bibitem[Vandenberghe and Boyd(2004)]{vandenberghe2004convex}
Lieven Vandenberghe and Stephen Boyd.
\newblock \emph{Convex optimization}, volume~1.
\newblock Cambridge university press Cambridge, 2004.

\bibitem[Wang et~al.(2022)Wang, Sankaran, and Perdikaris]{wang2022respecting}
Sifan Wang, Shyam Sankaran, and Paris Perdikaris.
\newblock Respecting causality is all you need for training physics-informed neural networks.
\newblock \emph{arXiv preprint arXiv:2203.07404}, 2022.

\bibitem[Wang et~al.(2023)Wang, Sankaran, Wang, and Perdikaris]{wang2023expert}
Sifan Wang, Shyam Sankaran, Hanwen Wang, and Paris Perdikaris.
\newblock An expert's guide to training physics-informed neural networks.
\newblock \emph{arXiv preprint arXiv:2308.08468}, 2023.

\end{thebibliography}

\newpage
\appendix

\section{Related Work}
We provide a review of the literature on neural HJ reachability and subsequent works that extend these methods to more complex and challenging problems. Additionally, we provide an overview of recent advances in methods for handling implicit Hamiltonians. 

\paragraph{High-dimensional neural HJ reachability.}
DeepReach \citep{bansal2021deepreach} introduced a mesh-free neural approach for solving HJ reachability problems in dimensions beyond the range of conventional grid-based solvers. Rather than discretizing the state space, DeepReach directly parameterizes the value function and minimizes the HJI variational inequality residual. Subsequent approaches have improved the scalability and training of neural HJ solvers. Sharpless et al.~\citep{sharpless2025linear} augment the HJI loss with supervision generated from tractable linearized differential games and demonstrate neural value-function learning for systems with tens of state dimensions. Similarly, Feng et al.~\citep{FengZ-RSS-25} and MADR~\citep{teoh2025madr} leverage MPC-based supervision to improve neural HJ reachability learning. These methods train the HJ value function end-to-end, but assume that the Hamiltonian required by the HJI residual can be evaluated efficiently. On the theoretical side, Hofgard~\citep{hofgard2024convergence} establishes conditional convergence guarantees for DeepReach, while Lin and Bansal~\citep{lin2023generating} provide formal safety assurances through error bounds and reachable-tube corrections. Related theoretical work also studies neural approximation of HJ-type PDEs~\citep{schaeffer2026neural}.

\paragraph{HJ reachability without closed-form Hamiltonians.}
Several recent works address HJ reachability problems in which the Hamiltonian is difficult to evaluate or unavailable in closed form. \citep{chilakamarri2025blackbox} learn a neural Hamiltonian surrogate from black-box dynamics samples for subsequent use in HJ solvers. 
\citep{choi2025ddh} construct a conservative data-driven Hamiltonian directly from trajectory data, while \citep{lin2026rover} extend HJ-based verification to complex perception-based controllers under state uncertainty. These approaches make a difficult Hamiltonian tractable through approximation or bounding before it is used by the value function solver. In contrast, we retain the numerical optimization defining the Hamiltonian as an explicit component of neural HJI training and differentiate the value-function objective through its solution.

\paragraph{Implicit Hamiltonians in Control.}
Implicit optimization layers have been studied broadly in machine learning and differentiable optimization, with Jacobian-Free Backpropagation (JFB), which provides an inexpensive alternative to exact implicit differentiation \citep{fung2022jfb, heaton2021feasibility, heaton2023explainable, mckenzie2024differentiating, mckenzie2024three}. In control, \citep{li2026end} combine operator splitting with JFB to enable scalable end-to-end training through embedded CBF-QP safety layers, which demonstrates that Jacobian-free differentiation can substantially reduce the computational cost of repeatedly differentiating through optimization problems along control trajectories. More closely related to our setting, \citep{gelphman2025end} develop an end-to-end method for high-dimensional optimal control in which the Hamiltonian minimization is solved implicitly and differentiated using JFB. Subsequent work establishes convergence guarantees for the resulting approximate-gradient training procedure \citep{gelphman2026convergence}. Imitation-based function-encoder policies instead sidestep the Hamiltonian, transferring zero-shot across tasks \citep{li2026zero}.
As shown in~\Cref{sec:theory}, the particular structure of the HJ reachability problem allows for exact gradient evaluation of the training objective without requiring approximate differentiation techniques, providing a more direct and natural alternative to existing approaches for handling implicit Hamiltonians.

\section{Additional Experimental Details}
\label{sec:example_details}

This appendix provides supplementary problem formulations, Hamiltonian derivations, and implementation details for the numerical experiments presented in~\Cref{sec:results}.

\subsection{High-Dimensional Energy-Constrained Pursuit--Evasion}

We extend the two-player pursuit--evasion problem in~\Cref{sec:results} to $N$ independent evader-pursuer pairs, with joint state $x=(r_1,q_1,\ldots,r_N,q_N)\in\mathcal{X}\subset\mathbb{R}^{2N}$. For each pair $i$, $r_i$ and $q_i$ denote the relative distance and energy resource, respectively, with controls $u_i\in[0,\bar u_i]$ and $d_i\in[0,\bar d_i]$.
The dynamics are
\begin{equation}
\begin{cases}
\dot r_i =
(1+k_i^1q_i)\tanh(u_i)
-1.2(1-k_i^2q_i)\tanh(d_i),\\
\dot q_i=-\alpha_i u_i^2+\beta_i d_i^2,
\end{cases}
\qquad i=1,\ldots,N,
\end{equation}
where $k_i^1,k_i^2,\alpha_i,\beta_i>0$ may vary across pairs to model heterogeneous maneuverability and energy consumption rate.
The target set consists of states in which at least one pair violates its separation or energy threshold, and is represented by the terminal function
\begin{equation}
\ell(x)=\min_{i=1,\ldots,N} \left\{ r_i-r_i^{\mathrm{reach}}, q_i-q_i^{\mathrm{reach}} \right\}.
\end{equation}
The corresponding target set is
$$
\mathcal{L} = \bigcup_{i=1}^{N} \left( \{r_i\le r_i^{\mathrm{reach}}\} \cup \{q_i\le q_i^{\mathrm{reach}}\} \right).
$$
Since the individual evader--pursuer pairs are dynamically decoupled, the Hamiltonian separates as $H(x,p)=\sum_{i=1}^{N}H_i(x_i,p_i)$, where $x_i=(r_i,q_i)$ and $p_i=[p_{r_i},p_{q_i}]^\top$.
Thus, the optimal control and disturbance can be computed independently for each pair using the same numerical optimization procedure as in the 2D case. 
We also note that under suitable  $k_i^1$ and $k_I^2$ we can preserve the convex-concave structure of the Hamiltonian, satisfying the necessary assumptions for the theory.

As the problem dimension increases, the reachable set occupies a larger portion of the state space, leading to an imbalance between reachable and unreachable states under uniform sampling. To address this imbalance and stabilize training, we employ importance sampling to ensure that decreasing training loss leads to improved closed-loop performance.

\begin{figure}[t]
    \centering
    \includegraphics[width=1.0\linewidth]{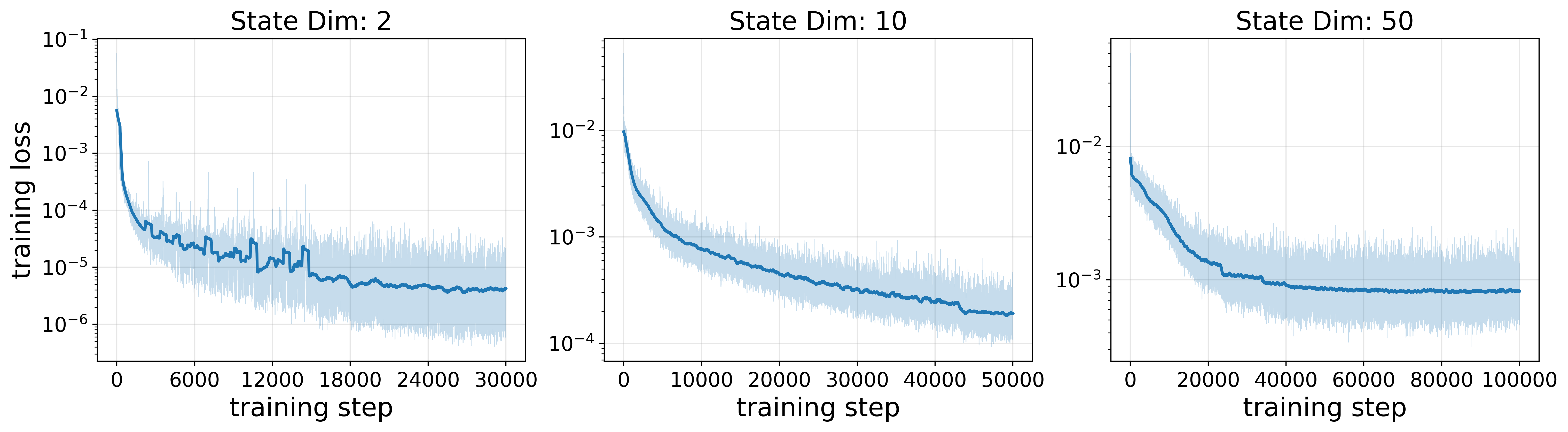}
    \caption{Training loss curves for the energy-constrained pursuit-evasion problems in 2D, 10D, and 50D. The losses decrease and stabilize across all three settings, demonstrating convergence of our proposed neural HJ training procedure. Higher-dimensional problems exhibit larger final residual losses.}
    \label{fig:scalability-results}
\end{figure}

\subsection{Two-Vehicle Collision Avoidance}

We consider a finite-horizon collision avoidance problem between two planar vehicles. The joint state is $x=(x_1,\ldots,x_6)\in\mathbb{R}^4 \times\mathbb{S}^1\times\mathbb{S}^1$, where $(x_1,x_2,x_3)$ and $(x_4,x_5,x_6)$ denote the positions and headings of the pursuer and evader, respectively. We modify the standard constant-speed dynamics~\citep{bansal2021deepreach} by introducing steering dependent speed for the evader:
\begin{equation}
\left\{
\begin{aligned}
\dot{x}_1 &= 0.75\cos x_3, &
\dot{x}_4 &= 0.75e^{-0.4(u/3)^2}\cos x_6,\\
\dot{x}_2 &= 0.75\sin x_3, &
\dot{x}_5 &= 0.75e^{-0.4(u/3)^2}\sin x_6,\\
\dot{x}_3 &= d, &
\dot{x}_6 &= u,
\end{aligned}
\right.
\end{equation}
where $u,d\in[-3,3]$ denote the evader's and pursuer's turn-rate controls, respectively. The evader's speed decreases with the magnitude of its turn rate, introducing nonlinear control dependence into the Hamiltonian.
For the time horizon $T=1$, the capture target is defined by
\begin{equation}
\ell(x)=
\sqrt{(x_1-x_4)^2+(x_2-x_5)^2}-0.25,
\label{eq:6d_target}
\end{equation}
with $\mathcal{L}=\{x:\ell(x)\leq0\}$. The corresponding Hamiltonian is
\begin{equation}
\begin{aligned}
H(x,p) ={}& 0.75(p_1\cos x_3+p_2\sin x_3)-3|p_3|\\
&+\max_{u\in[-3,3]} \Big[ 0.75e^{-0.4(u/3)^2} (p_4\cos x_6+p_5\sin x_6)+p_6u \Big],
\end{aligned}
\label{eq:6d_hamiltonian}
\end{equation}
where $p=\nabla_x V(x,t)$. While the pursuer's optimal turn rate admits a closed-form solution, the evader’s optimal control is determined through an inner optimization problem due to the nonlinear dependence of the Hamiltonian on its turn rate. 

We further observe that the curvature of the evader's maximization objective is determined entirely by the sign of the term $(p_4\cos x_6+p_5\sin x_6)$. Specifically, the objective is strongly concave for $p_4\cos x_6+p_5\sin x_6>0$ and strongly convex for $p_4\cos x_6+p_5\sin x_6<0$ over the admissible control set. Consequently, the global maximizer either lies on the domain boundary or can be provably recovered through methods such as projected gradient ascent.
Finally, we use the computational domain
\[
\mathcal{X} =[-1.5,1.5]^2\times[-\pi,\pi) \times[-1.5,1.5]^2\times[-\pi,\pi),
\]
with state-time samples drawn uniformly during training.

\begin{figure}
    \centering
    \includegraphics[width=0.95\linewidth]{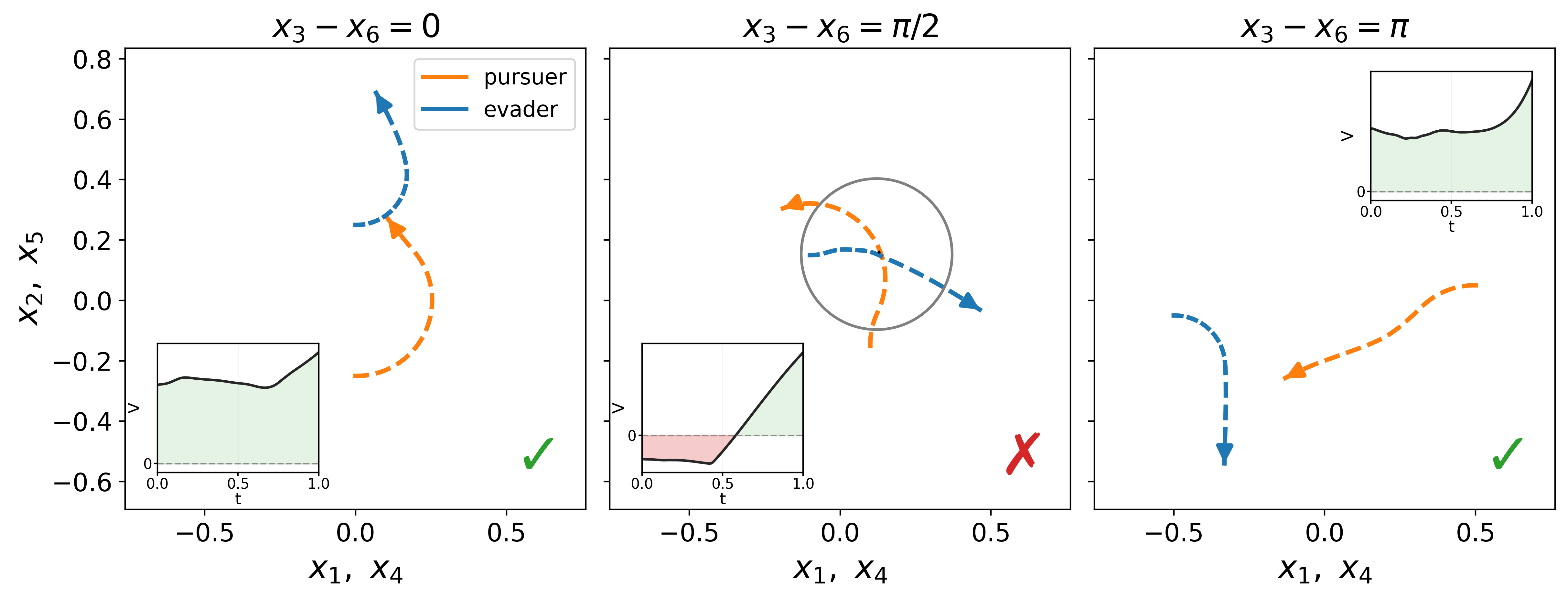}
    \caption{Closed-loop trajectories for the two-vehicle collision avoidance problem, using feedback controls obtained from the learned value function and online Hamiltonian optimization. \textbf{Left \& Right:} successful evasion; \textbf{Middle:} capture. The learned value function is overlaid along each trajectory to illustrate its temporal evolution.}
    \label{fig:two-vehicle-traj}
\end{figure}

\subsection{Quadrotor Reach-Avoid Control under Adversarial Disturbances}
We consider a six-dimensional planar quadrotor subject to adversarial wind disturbances, with dynamics and constraints specified below. 
Let $x=(x_1,x_2,x_3,x_4,x_5,x_6)\in\mathbb R^6$ denote the quadrotor's position, velocity, pitch angle and pitch rate, respectively. The two thrusters of the rotor can be controlled directly via $u\in U=[0,\bar u]^2$, the adversary chooses wind velocity $d\in D=\{d\in\mathbb R^2:\|d\|_2\le \bar d\}$. The quadrotor operates under standard dynamics as in~\citep{mahony2012multirotor,sewell2026neural}, which are given by
\begin{equation*}
    \begin{cases}
        \dot x_1 &= x_3, \\
        \dot x_2 &= x_4, \\
        \dot x_3 &= \displaystyle\frac{-(u_1+u_2)\sin x_5-k(x,d)(x_3-d_1)}{m},  \\
        \dot x_4 &= \displaystyle\frac{(u_1+u_2)\cos x_5-k(x,d)(x_4-d_2)}{m}-g_0,\\
        \dot x_5 &= x_6, \\
        \dot x_6 &= \displaystyle\frac{L}{I}(u_2-u_1).
    \end{cases}
\end{equation*}
Here $ m=1\,\mathrm{kg}$ is the mass, $I=0.05\,\mathrm{kg\,m^2}$ is pitch inertia, $L=0.2\,\mathrm m$ is the rotor's arm length, we use $g_0=9.81\,\mathrm{m/s^2}$ to denote gravity. We also set the maximum thrust to $ \bar u=7\,\mathrm N$ and the maximum wind speed to $\bar{d}=3\,\mathrm{m/s}$.
While the rotor dynamics are affine in control, the wind disturbance enters via a nonlinear drag force which we define here as
\[
 k(x,d)=c_1+c_2\sqrt{(x_3-d_1)^2+(x_4-d_2)^2}.
\]
Here the force model combines linear and quadratic drag based on relative air velocity, while not the highest fidelity aerodynamics model, it is still commonly used for simulation and analysis~\citep{hattenberger2023evaluation,kwon2024quadrotor}. For our example we consider coefficients $c_1 = 0.5$ and $c_2 = 0.05$. The terminal time is set to $T = 2 \mathrm s$. 
For simplicity, we consider a box region over which we sample and solve the resulting BRAT problem, which we define as 
$\Omega=[-2,2]\times[-0.5,4]\times[-3,3]\times[-3,3]\times[-1.2,1.2]\times[-4,4].$ 

The reach target of the problem is a boxed region over $x_1$ and $x_2$ at $[0,1]\times[2,3]$, we define the reach and avoid set functions as
\begin{equation*}
\begin{cases}
    \ell(x) &= \max\{-x_1, x_1-1, 2-x_2, x_2-3\}, \\
    g(x)    &= \max\left\{-x_2,\;|x_5|-\frac{\pi}{3}\right\},
\end{cases}
\end{equation*}
and subsequently the target and unsafe sets as
\[
\mathcal L=\{x:\ell(x)\leq 0\},\quad \mathcal G=\{x:g(x)>0\}
\]
following the conventions established in~\Cref{sec:background}. We note that the unsafe set encodes two failure scenarios: the first occurs when the quadrotor reaches the ground, and the second occurs when the pitch angle exceeds the prescribed safety limit, potentially leading to a loss of stability and a drone flip-over. We place no direct target restrictions on $x_3,x_4,x_5,x_6$,  although a successful arrival at the target must nevertheless satisfy $g(x)\leq 0$.
The optimal wind disturbance is determined by the constrained optimization problem
\begin{equation*}
    d^\star\in\arg\max_{\|d\|_2\le \bar d}\; -\frac{k(x,d)}{m}\Big[p_3(x_3-d_1)+p_4(x_4-d_2)\Big].
\end{equation*}
Under our choice of parameters, the maximization admits a unique solution for every state and adjoint, though obtaining $d^\star$ always requires an iterative solver, making this a challenging yet suitable testbed for our proposed approach.

\section{Additional Numerical Results}
\label{sec:additional_results}
\subsection{Numerical Verification of The Theory}
We provide numerical verification of our main theoretical results~\Cref{sec:theory} here in this section. We mainly consider the the 2D energy constrained pursuit–evasion problem as the example to focus on. 
Our focus is to validate the accuracy of the training gradient $\nabla_\theta \mathcal L$ under our proposed approach in saddle point optimization. For simplicity we apply $k$ steps of projected gradient ascent/descent with fixed step size of $0.1$. Early stopping is disabled to ensure a fair comparison. We compare two different scenarios:
\begin{itemize}
  \item \textbf{Gradient Free (ours):} the inner optimization is untracked and only the obtained saddle point is supplied to the loss function.
  \item \textbf{Automatic Differentiation (AD):} each inner optimization step is fully tracked and unrolled during backpropagation.
\end{itemize}
For value function construction we consider two different parameterizations: \textbf{(i)} a polynomial parameterization for the value function with sensitivity under control, and \textbf{(ii)} a neural network with the same size used in our numerical testing. For both cases we initialize the value functions so that the adjoints remain strictly positive,thus ensuring the min-max structure and a strictly interior saddle point. for each testing case, we sample with batch size $1024$ uniformly over the domain.

\begin{table}[t]
\small
  \centering
  \caption{Gradient consistency test under different value function parameterizations and solver accuracies. Here we define $\Delta \nabla_\theta \mathcal L := \nabla_\theta\mathcal{L}_{\text{ad}} - \nabla_\theta\mathcal{L}_{\text{ours}}$. All terms are in $2$-norm unless otherwise specified. Gradient norm for the polynomial case is $\| \nabla_\theta\mathcal L\| = 14.0$ and for the neural network case $\| \nabla_\theta\mathcal L\| = 6.7$. }
  \label{tab:gradient_check}
  \begin{tabular}{c|ccc|ccc}
    \toprule
      & \multicolumn{3}{c}{Polynomial $V_\theta$} & \multicolumn{3}{c}{Neural Network $V_\theta$} \\
    k  & $\| \Delta \nabla_\theta \mathcal L\|$ & $\| \partial_u \varphi \|_\infty$ & $\| \partial_d \varphi \|_\infty$ &  $\| \Delta \nabla_\theta \mathcal L\|$ & $\| \partial_u \varphi \|_\infty$ & $\| \partial_d \varphi \|_\infty$ \\
    \midrule
    1   & $2.4\times10^{-3}$  & $2.7\times10^{-1}$  & $1.2\times10^{-1}$ & $2.0\times10^{-4}$ & $4.7\times10^{-2}$ & $4.2\times10^{-2}$ \\
2   & $3.9\times10^{-3}$  & $2.1\times10^{-1}$  & $1.1\times10^{-1}$  & $3.9\times10^{-4}$ & $4.7\times10^{-2}$ & $4.2\times10^{-2}$ \\
5   & $5.2\times10^{-3}$  & $1.5\times10^{-1}$  & $6.8\times10^{-2}$  & $9.1\times10^{-4}$ & $4.5\times10^{-2}$ & $4.0\times10^{-2}$ \\
10  & $3.7\times10^{-3}$  & $7.8\times10^{-2}$  & $3.2\times10^{-2}$  & $1.6\times10^{-3}$ & $4.2\times10^{-2}$ & $3.8\times10^{-2}$ \\
20  & $1.0\times10^{-3}$  & $2.2\times10^{-2}$  & $7.3\times10^{-3}$  & $2.6\times10^{-3}$ & $3.6\times10^{-2}$ & $3.3\times10^{-2}$ \\
50  & $1.2\times10^{-5}$  & $4.4\times10^{-4}$  & $8.2\times10^{-5}$  & $3.3\times10^{-3}$ & $2.4\times10^{-2}$ & $2.3\times10^{-2}$ \\
100 & $8.7\times10^{-9}$  & $6.7\times10^{-7}$  & $4.6\times10^{-8}$  & $2.2\times10^{-3}$ & $1.3\times10^{-2}$ & $1.4\times10^{-2}$ \\
200 & $7.9\times10^{-15}$ & $1.6\times10^{-12}$ & $1.5\times10^{-14}$ & $5.5\times10^{-4}$ & $4.5\times10^{-3}$ & $5.4\times10^{-3}$ \\
500 & $0$                 & $1.1\times10^{-15}$ & $1.1\times10^{-15}$ & $7.4\times10^{-6}$ & $2.1\times10^{-4}$ & $3.1\times10^{-4}$ \\

    \bottomrule
  \end{tabular}
\end{table}

We report gradient consistency results for both test cases under different solver iterations $k \in \{1, 2, 5,10,\dots, 500 \}$ and display the results in~\Cref{tab:gradient_check}.
The gradient accuracy over the trainable weights  is governed by the accuracy of the inner solve, and vanishes with it.
As $k$ increases, the Hamiltonian residuals $\|\partial_u \varphi\|_\infty$ and $\|\partial_d \varphi\|_\infty$ decrease, and $\|\Delta\nabla_\theta\mathcal L\|$ decreases with them. 
For the polynomial $V_\theta$ the solve converges to machine precision by $k\approx200$, and the two gradients coincide exactly at $k=500$. The neural network converges more slowly because under random initialization its adjoint $p_\theta$ is small and the Hamiltonian is only weakly curved in the controls. Nevertheless the gradient discrepancies are negligible at $k=500$ for both cases especially regarding the $O(1)$ gradient norms. 
For cases where the saddle points lie on the boundaries, the results follow trivially. Specifically, when evaluated at the exact boundary, the gradient difference is exactly $0$, since no fixed-point error needs to be accounted for. Overall the numerical results fully verifies our theoretical claims.

\subsection{Inner Solver Comparison}
The inner saddle point problem~\eqref{eq:hamiltonian} is solved approximately at every training step. While our theory in~\Cref{sec:theory} ensures gradient consistency at the fixed point, the quality of the truncated solve can still affect training.
We hereby conduct an ablation study and compare between fixed and adaptive step size selection in the saddle point optimization under the same gradient descent rule. We test on the 10D energy constrained pursuit–evasion example, where both optimizers are given the same budget of $200$ iterations. Two copies of the value network are trained under exactly the same sampling and with identical weight initialization; hence, any difference in training is due to the inner solver alone. 
We also keep a reference solution for the fixed point solver, which uses PGDA with a fixed step size of $0.1$ and $5000$ iterations. This achieves a median fixed point error of $1.6\times10^{-6}$ at each sampling mini-batch and serves as a ground truth reference when comparing gradient accuracies. 
We run a total of $10000$ training steps for the test. 

\begin{table}[t]
\small
\centering
\caption{Inner solver accuracy comparison between fixed and adaptive step size selection. Hamiltonian gap and gradient alignment measure relative value and gradient errors against the reference solution. Numbers averaged across 10K training steps except for the final loss.}
\label{tab:inner_solver}
\begin{tabular}{cccccc}
\toprule
 & \makecell{\% of stationary \\ samples$^\ast$} & \makecell{Hamiltonian \\ gap $\downarrow$} & \makecell{Gradient \\alignment $\uparrow$}  & \makecell{Final Training \\ loss $\downarrow$}  & \makecell{Per training step \\ solver time $\downarrow$} \\
\midrule
Fixed $\eta = 0.1$     & $22\%$ & $4.8\%$ & $0.965$  & $8.55\times 10^{-4}$   & $\mathbf{0.214}$s \\
Adaptive $\eta$   & $\mathbf{98.9\%}$ & $\mathbf{0.4\%}$ & $\mathbf{0.999}$  &  $\mathbf{8.30\times 10^{-4}}$  & $0.247$s \\
\bottomrule
\end{tabular}

\vspace{2pt}
\footnotesize
$^\ast$ we use tolerance at $1 \times 10^{-3}$ to determine convergence.
\end{table}

We present the results in~\Cref{tab:inner_solver}. Under the same iteration budget, adaptive step sizing yields a substantially more accurate inner solve than a carefully selected fixed step size. Importantly, the computational overhead is marginal, allowing us to maintain the fast training required for neural HJ reachability at large.
However, we note that despite the lower solver accuracy, using plain PGDA with a fixed step size does not lead to significant errors in either the gradient updates or the overall training loss behavior. We attribute this to the stochasticity of Adam and mini-batch training in general. The key takeaway we want to highlight is that convergence of the value function can still be expected even when the inner solve admits a small error.

\section{Training Setup and Details on Numerical Experiments}
\label{sec:training_details}

We provide details on the training setup used in this paper. We consider a total of five different experiments.
In all examples the value function $V_\theta$ is defined per~\eqref{eq:hard_terminal}, with $N_\theta$ being a residual MLP with $\tanh$ activations and terminal condition enforced exactly by construction. Training only minimizes the VI residual at sampled locations.
For all examples except for the quadrotor, we employ a uniform sampling scheme over the state-time space.
We train all networks with Adam optimizer~\citep{kingma2014adam} with an initial learning rate of $10^{-3}$. The learning rate is halved whenever training hits a plateau, the patience for which we set differently for each example.
The default solver for the inner optimization is PGDA with Barzilai--Borwein adaptive step sizing, as introduced in~\Cref{sec:method} with initial step size $0.1$. We terminate early whenever the fixed-point residual reaches $1\times 10^{-4}$. The maximum number of iterations for the inner optimization is set separately for each individual example. We try to balance between solver accuracy and speed during training.
All code is implemented in PyTorch~\citep{paszke2019pytorch}, and all runs use random seed $0$; problem-specific settings are listed in Table~\ref{tab:training_hyperparameters}.

\begin{table}[t]
\small
\centering
\caption{Problem-specific training settings. Inner iterations are the maximum number of iterations of the inner solver.}
\label{tab:training_hyperparameters}
\begin{tabular}{ccccccc}
\toprule
Example & $T$ & \makecell{Training \\Steps} & \makecell{Batch \\ Size} & \makecell{Network Size \\ (Width, \# of Layers)} & \makecell{LR \\ patience} & \makecell{Inner \\Iters.} \\
\midrule
\makecell{Energy Constrained \\ Pursuit–Evasion (2D)} & 1.5 & 30k  & 256  & (128, 3) & 2000 & 100 \\
\makecell{Energy Constrained \\ Pursuit–Evasion (10D)} & 1.5 & 50k  & 256  & (128, 4) & 3000 & 200 \\
\makecell{Energy Constrained \\ Pursuit–Evasion (50D)} & 1.5 & 100k & 256  & (256, 4) & 5000 & 500 \\
\makecell{Two-Vehicle \\ Collision Avoidance} & 1.0 & 50k & 1024 & (128, 4) & 3000 & 100 \\
\makecell{Planar Quadrotor \\ (BRAT)}& 2.0 & 70k & 1024 & (256, 4) & 3000 & 10$^\ast$ \\
\bottomrule
\end{tabular}

\vspace{2pt}
\footnotesize
$^\ast$ We use a problem-specific solver for the BRAT example based on global contraction, which converges faster and is more accurate than PGDA.
\end{table}

\section{Proof of Theorem}
\label{sec:proof}
For completeness, we restate the assumptions and main theorem from~\Cref{sec:theory} and provide its detailed proof, together with the supporting lemmas.

\paragraph{Notations.} Write $z=(u, d)\in \U \times \D $. We denote
\begin{equation}\label{eq:p_abbrv}
    p_\theta:=\nabla_x V_\theta(x,t),
\end{equation}
\begin{equation}\label{eq:varphi}
    \varphi(p,u,d)=\varphi(p,z):=p^\top f(x,u,d)=p^\top f(x,z),
\end{equation}
where we slightly abuse notations, and used $z$ and $(u,d)$ interchangeably,
and thus
\begin{equation*}
    H(x,p)=\max_{u\in U}\,\min_{d\in D}\, p^\top f(x,u,d)=\max_{u\in U}\,\min_{d\in D}\,\varphi(p,u,d).
\end{equation*}
\paragraph{Assumptions.} 
For an arbitrary parameter $\theta_0$, we assume the following:
\begin{enumerate}[label=\textbf{(A\arabic*)}, leftmargin=*, itemsep=4pt]
\item $\U$ and $\D$ are nonempty, compact and convex; 
\item $f(x,\cdot)$ is $C^1$;
\item the mapping $\theta\mapsto p_\theta$ is differentiable at $\theta_0$; 
\item there exists $\delta_0=\delta(p_{\theta_0})>0$ such that, for every
$p \in \mathcal{B}_0 := \{ p \in \mathbb{R}^n: \norm{p-p_{\theta_0}}<\delta_0 \}$, the Isaacs condition
\[
  \max_{u\in\mathcal{U}}\min_{d\in\mathcal{D}}\varphi(p,u,d)
  =\min_{d\in\mathcal{D}}\max_{u\in\mathcal{U}}\varphi(p,u,d)
\]
holds; 
\item the function $\varphi(p_{\theta_0},\cdot,\cdot)$ has a unique saddle point on
$\mathcal{U}\times\mathcal{D}$, which we denote by $z^\star(p_{\theta_0})$.
\end{enumerate}

\mainthm*

\noindent\textbf{Proof sketch: } The key idea of our proof is to avoid differentiating through the saddle point solution map, which may be nondifferentiable. Instead, we directly analyze perturbations of the Hamiltonian using saddle point inequalities. Together with compactness and uniqueness, these bounds establish differentiability of the Hamiltonian and yield the desired gradient identity through the chain rule.

\newcommand{\Sequation}{\mathcal{S}(p) := \left\{ z=(u,d)\in \mathcal{U}\times \mathcal{D}: \; \varphi(p,\tilde{u},d) \leq \varphi(p,u,d) \leq \varphi(p,u,\tilde{d})\quad \forall \tilde{u} \in \mathcal{U},\, \tilde{d} \in \mathcal{D}  \right\}.}
\begin{proof}
    Let $S(p)$ denote the set of saddle points at $p$, that is,
    \begin{equation}\label{eq:Sequation}
        \Sequation
    \end{equation}

    \paragraph{Step 1: Existence of Saddle Point.}
    By (A4), for every
$p \in \mathcal{B}_0 := \{ p \in \mathbb{R}^n: \norm{p-p_{\theta_0}}<\delta_0 \}$, the Isaacs condition
    \begin{equation}\label{eq:def_v}
    \max_{u\in\mathcal{U}}\,\min_{d\in\mathcal{D}}\,\varphi(p,u,d)=\min_{d\in\mathcal{D}}\,\max_{u\in\mathcal{U}}\,\varphi(p,u,d)=:v
    \end{equation} 
    holds. 
    Here, the extrema are attained by compactness and continuity. Let $u^\star$ attain the outer maximum on the left and $d^\star$ the outer minimum on the right. On one hand,
    \[
    \varphi(p,u^\star,d^\star) \geq \min_{d\in\mathcal{D}}\varphi(p,u^\star,d) = v,
    \]
    where in the first step, we used the property of minimum, and in the second step, we used the definition of $v$ in~\eqref{eq:def_v}. On the other hand, 
    \[
    \varphi(p,u^\star,d^\star) \leq \max_{u\in\mathcal{U}}\varphi(p,u,d^\star)=v,
    \]
    using the same argument. Combining these two inequalities, we have
    \begin{equation}\label{eq:varphi_ustar_dstar}
        \varphi(p,u^\star,d^\star)=v.
    \end{equation}
    Next, we argue that $(u^\star, d^\star)$ is a saddle point. On one hand, we have, for every $\tilde u\in\mathcal{U}$,
\begin{equation}\label{eq:l_saddle}
\varphi(p,\tilde u,d^\star)\ \le\ \max_{u\in\mathcal{U}}\varphi(p,u,d^\star)\ =\ v\ =\ \varphi(p,u^\star,d^\star),
\end{equation}
by the definition of $d^\star$. On the other hand, similarly, for every $\tilde d\in\mathcal{D}$,
\begin{equation}\label{eq:r_saddle}
\varphi(p,u^\star,\tilde d)\ \ge\ \min_{d\in\mathcal{D}}\varphi(p,u^\star,d)\ =\ v\ =\ \varphi(p,u^\star,d^\star),
\end{equation}
by the definition of $u^\star$. Together, \eqref{eq:l_saddle} and \eqref{eq:r_saddle} establish that $(u^\star, d^\star)$ is a saddle point, that is $(u^\star, d^\star) \in \mathcal{S}(p)$. Thus, $\mathcal{S}(p) \neq \emptyset$ for every $p \in \mathcal{B}_0$, which is the existence of saddle point.
Moreover, by the definition of the Hamiltonian $H$, \eqref{eq:varphi_ustar_dstar} becomes
\begin{equation}\label{eq:Ham_saddle}
    H(x,p)=\varphi(p,u^\star, d^\star).
\end{equation}

\paragraph{Step 2: Linearization Error Bound.} We bound the error of linearizing $H$ around $p_{\theta_0}$.
Given any fixed $p' \in \mathcal{B}_0$, write $(u_0,d_0):= z^\star(p_{\theta_0})$ and let $(u',d') \in \mathcal{S}(p')$ be any saddle point at $p'$ (which exists by Step~1). 

\noindent\textit{Upper Bound.} Using~\eqref{eq:SV} of~\Cref{lemma:SV}, we have
\begin{align*}
    H(x, p') &\leq \varphi(p', u', d_0), \\ 
    H(x, p_{\theta_0}) &\geq \varphi(p_{\theta_0}, u', d_0).
\end{align*}
Subtracting these two inequalities, we have 
\begin{equation}\label{eq:upper_SV}
    H(x, p') - H(x, p_{\theta_0}) \leq (p' - p_{\theta_0})^\top f(x, u', d_0).
\end{equation}
\textit{Lower Bound.} Similarly, using~\eqref{eq:SV} of~\Cref{lemma:SV}, we have
\begin{align*}
    H(x, p') &\geq \varphi(p', u_0, d'), \\ 
    H(x, p_{\theta_0}) &\leq \varphi(p_{\theta_0}, u_0, d').
\end{align*}
Subtracting these two inequalities, we have
\begin{equation}\label{eq:lower_SV}
    H(x, p') - H(x, p_{\theta_0}) \geq (p' - p_{\theta_0})^\top f(x, u_0, d').
\end{equation}
\textit{Error Bound.} Let the linearization error be
\begin{equation}\label{eq:remain_SV}
    R(p') := H(x, p') - H(x,p_{\theta_0}) - (p'-p_{\theta_0})^\top f(x, u_0, d_0).
\end{equation}
Plugging~\eqref{eq:upper_SV} and \eqref{eq:lower_SV} into \eqref{eq:remain_SV}, we have 
\begin{equation}\label{eq:remain_abs_bound}
    (p'-p_{\theta_0})^\top(f(x,u_0,d')-f(x,u_0,d_0)) \leq R(p') \leq (p'-p_{\theta_0})^\top(f(x,u',d_0)-f(x,u_0,d_0)).
\end{equation}
Next, we establish the linearization error bound as
\begin{equation}
    \left| H(x, p') - H(x,p_{\theta_0}) - (p'-p_{\theta_0})^\top f(x, z^\star(p_{\theta_0})) \right| = |R(p')|
\end{equation}
by the definition \eqref{eq:remain_SV}, and $z^\star(p_{\theta_0})=(u_0,d_0)$, and 
\begin{equation}
    |R(p')| \leq \norm{p' - p_{\theta_0}} \omega(p') \label{eq:R_bound}
\end{equation}
by applying the Cauchy--Schwarz inequality to~\eqref{eq:remain_abs_bound}, where
\begin{equation}
    \omega(p') = \max \Big\{ \norm{f(x,u_0,d')-f(x,u_0,d_0)}, \norm{f(x,u',d_0)-f(x,u_0,d_0)} \Big\}.
\end{equation}

\paragraph{Step 3: Differentiability of $H$.} We prove that $H(x, \cdot)$ is differentiable at $p_{\theta_0}$. 
By~\Cref{lemma:cts_saddle}, as $\mathcal{B}_0 \owns p' \to p_{\theta_0}$, then for any choice of saddle point $(u',d') \in \mathcal{S}(p')$ (which exists by Step~1), $(u',d') \to (u_0, d_0)$. By (A2), $f(x, \cdot)$ is continuous, thus $\omega(p') \to 0$ as $p' \to p_{\theta_0}$. Therefore, 
\[
|R(p')| / \norm{p' - p_{\theta_0}} \leq \omega(p') \to 0
\]
as $p' \to p_{\theta_0}$. Since $R(p')$ does not depend on the choice of the saddle point $(u',d')$, this implies that $H(x, \cdot)$ is differentiable at $p_{\theta_0}$, and $\nabla_p H(x, p) \big|_{p_{\theta_0}} = f(x, z^\star(p_{\theta_0}))$. 

\paragraph{Step 4: Gradient-free.} By (A3), $p_\theta$ is differentiable at $\theta_0$, thus by Step~3 and the chain rule,
\[
    \partial_\theta H(x,p_\theta)|_{\theta_0} = \left(\nabla_p H(x, p) \big|_{p_{\theta_0}}\right)^\top (\partial_\theta p_\theta) |_{\theta_0} =  f(x, z^\star(p_{\theta_0}))^\top \, (\partial_\theta p_\theta) |_{\theta_0}.
\]

\end{proof}

\paragraph{Supporting Lemmas.}
\begin{lemma}[Saddle Value] \citep[Section~5.4.2]{vandenberghe2004convex} \label{lemma:SV}
If $(u^\star,d^\star)$ is a saddle point at $p$, then $H(x,p)=\varphi(p,u^\star,d^\star)$ and
\begin{equation}\label{eq:SV}
\varphi(p, u,d^\star)\ \le\ H(x,p)\ \le\ \varphi(p,u^\star, d),
\qquad\forall\, u\in\U,\  d\in\D. \tag{SV}
\end{equation}
\end{lemma}

\begin{lemma}[Continuity of saddle points at $p_{\theta_0}$]\label{lemma:cts_saddle} Let $S(p)$ denote the set of saddle points at $p$, that is,
\begin{equation}\tag{\ref*{eq:Sequation}}   
    \Sequation
\end{equation}
Under (A1), (A2), and (A5), if the sequence $p_k \to p_{\theta_0}$ with $p_k \in \mathcal{B}_0$ (defined in (A4)) and $z_k \in \mathcal{S}(p_k)$, then $z_k \to z^\star(p_{\theta_0})$.
\end{lemma}

\begin{proof}
    Write $z_0:=z^\star(p_{\theta_0})$ and $z_k=(u_k, d_k)$ for all $k$. Suppose the contrary, that is, there exists $\epsilon > 0$ such that there is a subsequence $\{ z_{k'} \}$ with $\norm{z_{k'} - z_0} \geq \epsilon \,$ for all $k'$. 
    
    By (A1), since $\mathcal{U} \times \mathcal{D}$ is compact, by the Bolzano-Weierstrass theorem~\cite[Theorem 3.4.8]{bartle2000introduction}, there exists a convergent subsequence $\{ z_{k''} \}$ of $\{ z_{k'} \}$ such that $z_{k''} \to \bar{z}=(\bar{u}, \bar{d}) \in \mathcal{U} \times \mathcal{D}$ and 
    \begin{equation}\label{eq:bar_z_contrary}
        \norm{\bar{z}-z_0} \geq \epsilon,
    \end{equation} by continuity of $\norm{\cdot}$. Also, remark that $p_{k''} \to p_{\theta_0}$ by assumption.
    Since $z_{k''}=(u_{k''}, d_{k''}) \in \mathcal{S}(p_{k''})$, we have~\citep[Section~5.4.2]{vandenberghe2004convex}
    \begin{equation}\label{eq:SV_subsubseq}
    \varphi(p_{k''}, \tilde{u}, d_{k''}) \leq \varphi(p_{k''}, u_{k''}, d_{k''}) \leq \varphi(p_{k''}, u_{k''}, \tilde{d}), \qquad\forall\, \tilde{u}\in\U,\  \tilde{d}\in\D.
    \end{equation}
    Consider $\varphi(p, u, d) = p^\top f(x, u, d)$. By (A2), $f(x, \cdot)$ is continuous on $\mathcal{U} \times \mathcal{D}$. Together with continuity of inner product, they imply $\varphi$ is jointly continuous in $(p, u, d)$. Thus, taking $k'' \to \infty$ gives
    \[
    \varphi(p_{\theta_0}, \tilde{u}, \bar{d}) \leq \varphi(p_{\theta_0}, \bar{u}, \bar{d}) \leq \varphi(p_{\theta_0}, \bar{u}, \tilde{d}),  \qquad\forall\, \tilde{u}\in\U,\  \tilde{d}\in\D.
    \]
    This implies that $\bar{z}=(\bar{u}, \bar{d})$ is a saddle point at $p_{\theta_0}$~\citep[Section~5.4.2]{vandenberghe2004convex}. On the other hand, by (A5), $\mathcal{S}(p_{\theta_0})=\{ z_0\}$ is a singleton. Thus, we have $\bar{z} \in \mathcal{S}(p_{\theta_0})=\{ z_0\}$, but this contradicts~\eqref{eq:bar_z_contrary} and thus contradicts our original assumption.
\end{proof}

\end{document}